\documentclass[]{revtex4}

\usepackage{graphicx}
\usepackage{amssymb}
\usepackage{subfigure}
\usepackage{epstopdf}
\newtheorem{proposition}{Proposition}

\begin{document}

\title{Actin automata: Phenomenology and localizations}
\author{Andrew Adamatzky and Richard Mayne}
\affiliation{Unconventional Computing Center, University of the West of England, Bristol BS16 1QY, UK}


\date{\today}


\begin{abstract}
Actin is a globular protein which forms long filaments in the eukaryotic cytoskeleton, whose roles in cell function include structural support, contractile activity to intracellular signalling. We model actin filaments as two chains of one-dimensional binary-state semi-totalistic automaton arrays to describe hypothetical signalling events therein. Each node of the actin automaton takes state `0' (resting) or `1' (excited) and updates its state in discrete time depending on its neighbour's states. We analyse the complete rule space of actin automata using integral characteristics of space-time configurations generated by these rules and compute state transition rules that support travelling and mobile localizations. Approaches towards selection of the localisation supporting rules using the global characteristics are outlined. We find that some properties of actin automata rules may be predicted using Shannon entropy, activity and incoherence of excitation between the polymer chains. We also show that it is possible to infer whether a given rule supports travelling or stationary localizations by looking at ratios of excited neighbours are essential for generations of the localizations. We conclude by applying biomolecular hypotheses to this model and discuss the significance of our findings in context with cell signalling and emergent behaviour in cellular computation.
\end{abstract}


\maketitle

\section{Introduction}

\begin{figure}[!bp]
\centering
\includegraphics[width=0.5\textwidth]{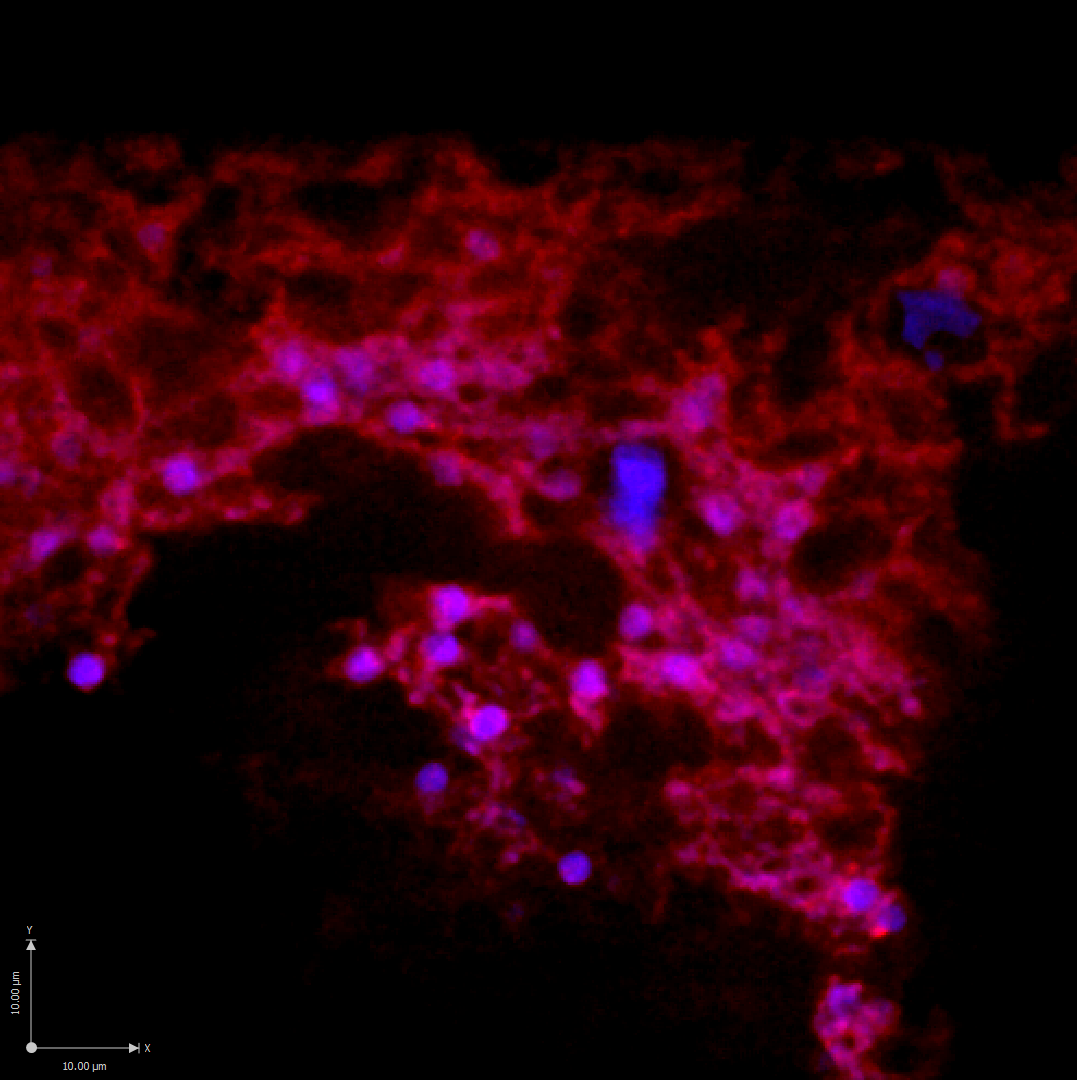}
\caption{Confocal laser scanning micrograph of the intracellular actin network (red) of slime mould \emph{Physarum polycephalum}. Articulations onto the slime mould's myriad nuclei (blue) can be seen.}
\label{actinphoto}
\end{figure}

Actin is a highly abundant protein which is expressed in all eukaryotic cells. A major component of the cellular cytoskeleton (Fig.~\ref{actinphoto}), it forms an intracellular scaffold whose functions are multifarious and completely essential to life. These include maintaining the cell's structural integrity, contributing to muscle contraction (via integral tropomyosin units) and trafficking cytoplasmic molecules and organelles \cite{Dominguez_Holmes_2011}. The roles of the actin component cytoskeleton are concomitant with those of the entire cytoskeleton, which is comprised of helical actin microfilaments, bundles of tubulin microtubules and a range of intermediate filaments and cytoskeletal-binding proteins. Where actin and tubulin are ubiquitous and highly conserved, the range of intermediate filaments present in a cell differ to represent its role, e.g. human epithelia contain large amounts of cytokeratins for enhanced structural integrity \cite{Southgate_etal_1999}.

The cytoskeleton is also thought to be extensively involved in cell signalling events. These may include, but are not limited to, signal transduction pathways, electrical potential, quantum events (e.g. changes in protein conformational state) and mechanical stress \cite{Hameroff_1998,Janmey_1998,Jibu_etal_1994}. When viewed from a perspective of natural computation, the cytoskeleton represents an attractive model for describing how data is transduced and transmitted within a cell. Whilst exact mechanisms for many of these process have yet to be fully elucidated, it has been suggested that `data' travelling through actin filaments may undergo computation via collisions between mobile localizations 
representing the 
data~\cite{Adamatzky:2001:NMC:644307.644322,zhang2009collision,adamatzky2011topics,adamatzky1997universal,adamatzky2000collision,toth2009simple}. The cytoskeleton may also play the role of a cell's `processor' to some degree, e.g. by Boolean logical operations as signals pass through certain proteins or branches in the cytoskeletal network~\cite{Lahoz-Beltra_etal_1993}. 

Cytoskeletal actin is a filamentous protein (f-actin) arranged in a double helix, comprised of polymerised globular (g-)actin monomers. Microfilaments are typically less than $1\mu$m in length and 8nm (80\r{A}) in diameter \cite{Mofrad_Kamm_2006}. Individual actin filaments are polymerised, joined together, crosslinked and bound to the cytoskeleton and other cellular structures by a range of actin-bnding proteins (ABPs); examples of common ABPs which fulfil these functions are profilin, Arp2/3 complex, filamin, spectrin and $\alpha$-actinin, respectively \cite{Winder_Ayscough_2005}.

\begin{figure}
\centering
\includegraphics[width=0.8\textwidth]{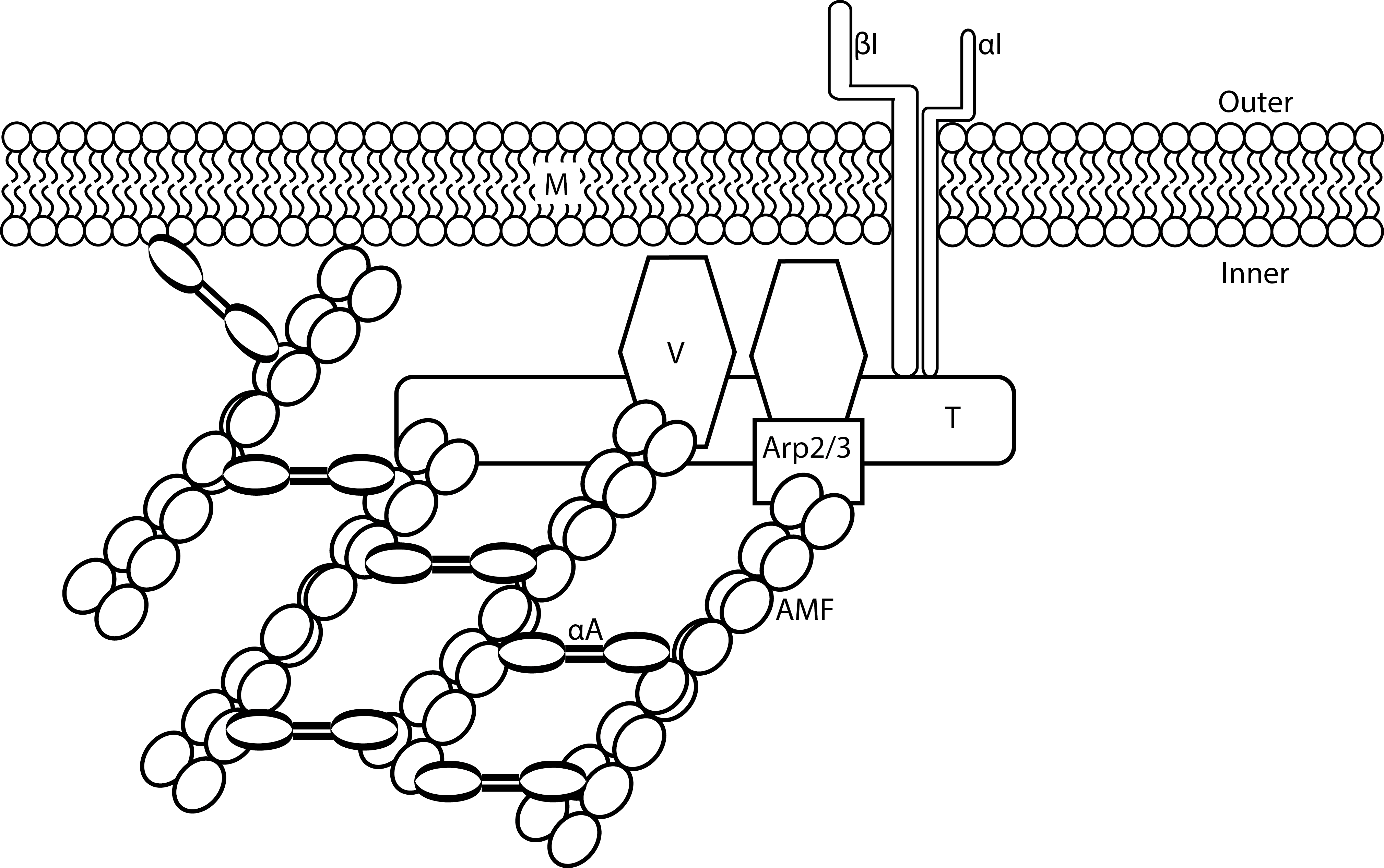}
\label{fig-fac}
\caption{Simplified schematic diagram showing the component proteins of the focal adhesion complex. Actin is able to detect mechanical stress and extracellular signalling events from the membrane/stimulation of integrins bound to the extracellular matrix. Signalling pathways are not shown. $\alpha$-I \& $\beta$-I: alpha \& beta integrin; M: phospholipid membrane; V: vinculin; T: talin; Arp2/3: Arp2/3 protein complex; AMF: actin microfilament; $\alpha$A: alpha actinin. Adapted from \cite{Janmey_1998,Mofrad_Kamm_2006,Petit_Thiery_2000}. }
\end{figure}

As a signalling element, much evidence is available to demonstrate how actin is able to detect and transduce a range of environmental stimuli by its associations with the cell membrane and membrane-bound proteins, such as ion channels and receptors \cite{Janmey_1998}. For example, cell-cell signalling events and mechanical stress may be transmitted by actin filaments via the focal adhesion complex; a complex multi-protein structure bound to the cell membrane Fig.~\ref{fig-fac}, the stimulation of which causes the transduction of mechanical `signals' down associated microfilaments via a variety of signalling cascades and/or direct transmission of mechanical force \cite{Hwang_Barakat_2012,Petit_Thiery_2000}. 

There is also evidence to suggest that actin may also transmit electrical potential/ionic waves \cite{Tuszynski_etal_2004} and quantum protein transitions \cite{Belmont_etal_1999,Hameroff_1998}, in addition to mechanical force and signalling cascades. These events may therefore also be considered as discrete data packets from a computational perspective.

In a previous study, we have discussed the putative role of the cytoskeleton in facilitating apparently intelligent behaviour in non-animal models (Fig. \ref{actinphoto}) via structuring of cellular sensorimotor data streams \cite{Mayne_etal_2014}. In this investigation, we model a generalised `information transmission' energetic event which propagates from one G-actin molecule to its neighbours through the chemical bonds between actin molecules. When stimulated, a G-actin molecule enters an excited state which corresponds with a net increase in energy; when this state is transferred to neighbouring molecules, the molecule returns to a non-excited resting state. This model has been engineered to apply to multiple forms of `data' which may be transmitted in the acin-component cytoskeleton; e.g. conduction of electrical potential, quantum energy states or physical waves/strand compression. The model intimates that actin filaments are stimulated at a hypothetical point where environmental data is perceived by the cell --- i.e. articulations to the cell membrane or membrane-bound receptors --- and that the associated `output' is a generalised system by which the data packet elicits a repeatable, predictable response, e.g. activation of a signal transduction cascade at a target protein within an organelle.

\section{Automaton model}

  \begin{figure}[!tbp]
 \centering
 \subfigure[]{\includegraphics[width=0.30\textwidth]{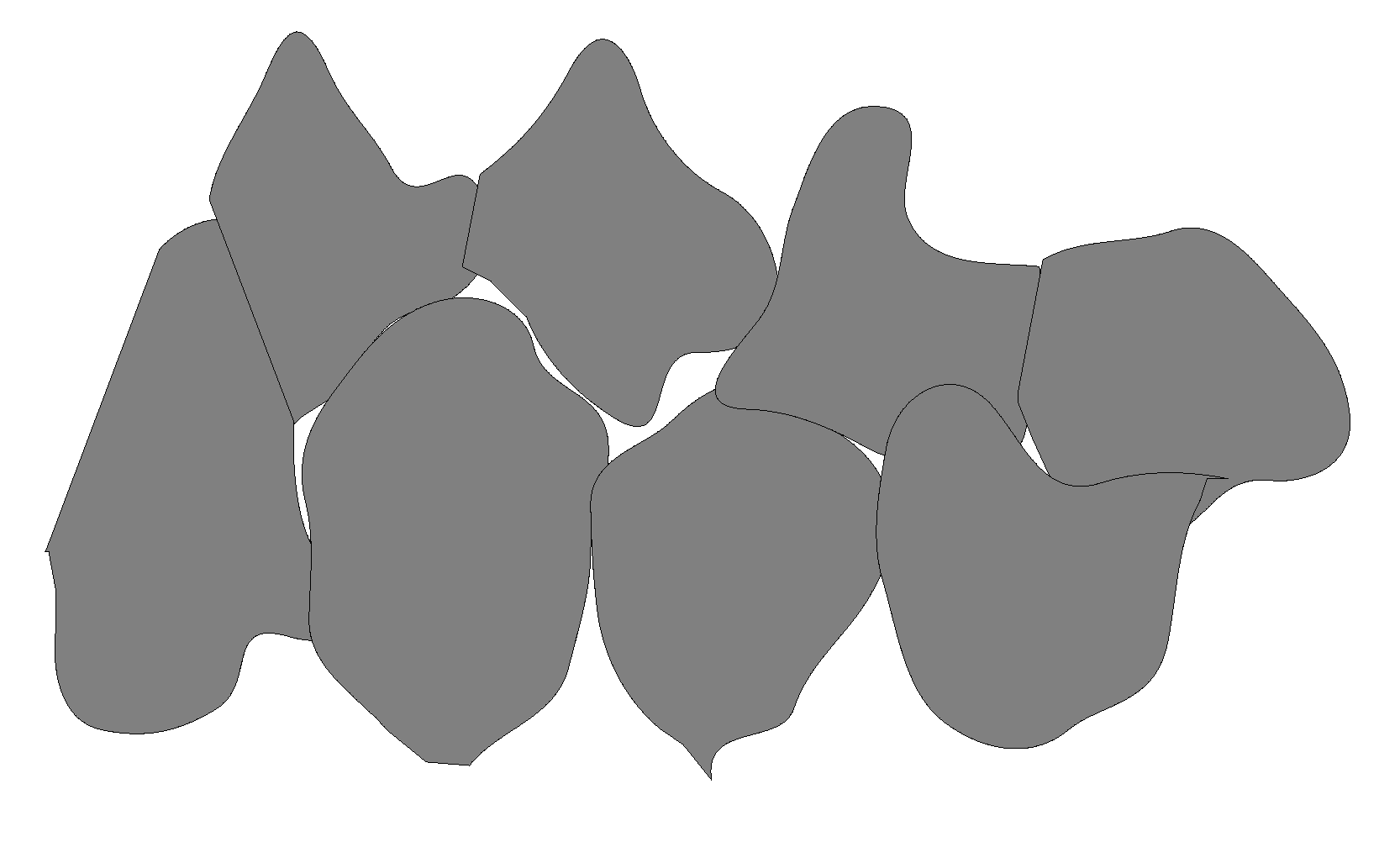}}
 \subfigure[]{\includegraphics[width=0.49\textwidth]{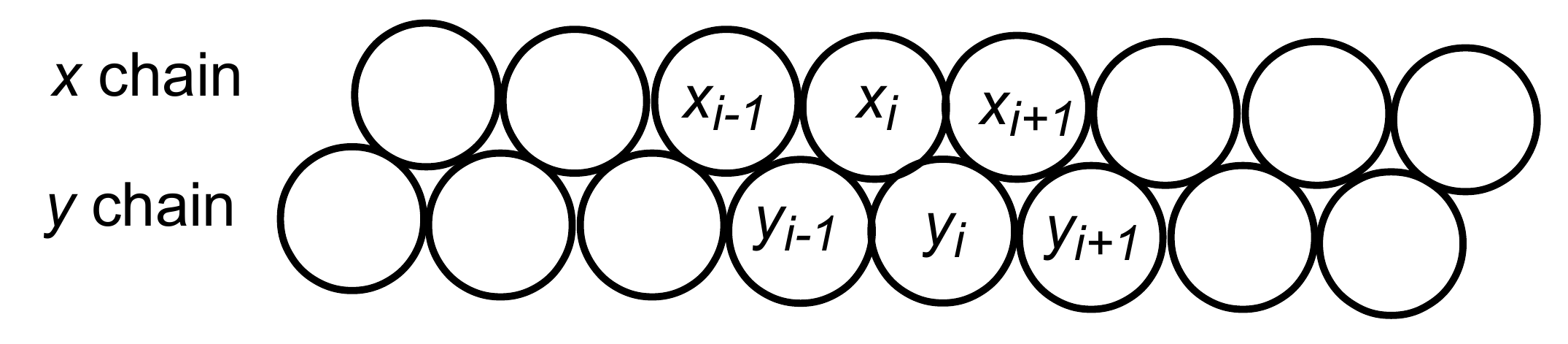}}
\caption{Schemematic digram of F-actin strands. (a)~Structure of actin detected by X-ray fiber diffraction. Adapted from~\cite{oda2009nature}. (b)~Actin automata.}
 \label{scheme}
 \end{figure}
 
 Each G-actin molecule (except those at the ends of F-actin strands) has four neighbours, as demonstrated 
 in Fig.~\ref{scheme}. An actin automaton consists of two chains $x$ and $y$ of semi-totalistic
  binary-state automata. Each automaton takes two states `0' (resting) and `1' (excited). 
 Automata update their states by the same rule in discrete time.
Each automaton updates its state depending on states of its immediate neighbours.
 The neighbourhood of an  automaton $x_i$ in chain $x$ (Fig. \ref{scheme}) is $u(x_i)=\{ x_{i-1}, x_{i+1}, y_{i}, y_{i-1} \}$ and 
 neighbourhood of an  automaton $y_i$ in chain $y$ is $u(y_i)=\{ y_{i-1}, y_{i+1}, x_{i}, x_{i+1} \}$ 
 states of automata $x_i$ and $y_i$ at time step $t$ are  $x^t_i$ and $y^t_i$.
 
Let  $\sigma^t_{x_i}=x^t_{i-1}+x^t_{i+1}+y^t_{i}+y^t_{i-1}$ and 
$\sigma^t_{y_i}=y^t_{i-1}+y^t_{i+1}+x^t_{i}+x^t_{i+1}$ be 
sums of excited, state `1', neighbours of automata $x_i$ and $y_i$. Automaton state transition function is determined by  a two-dimensional  matrix $F=(f_{i,j})$, $0 \leq i \leq 1$ and $0 \leq j \leq 4$, $f_{i,j} \in \{ 0, 1 \}$.  
Each automaton updates its states as the following  $x^{t+1} = f_{x^t_i,\sigma^t_{x_i}}$ and $y^{t+1} = f_{y^t_i,\sigma^t_{y_i}}$. There are 1024 states of $F$: each determines a unique automaton state transition rules. 
We will decode the rules via decimal representations of sub-matrices
 $F_0$ and $F_1$. For example, rule $(10,4)$ encodes the transition when $F_0=(01010)$ and $F_1=(00100)$, or in words,   a node  in state `0' takes state `1', or excites, if it has one or three neighbours in state `1' (otherwise, the cell remains in the resting state '0') and a cell in state `1' remains in the  state `1', or remains excited, if it has two neighbours in state `1'; otherwise, the cells switches to state '0', or returns to the resting state.
  
Assuming there are $n$ automata in each chain and the actin automaton evolves for $\tau$ time step, the following integral measures are calculated on space-time configurations generated by actin automata rules. In experiments presented here 
$n=300$ and $\tau=1000$. To compute the characteristics defined below we excited a resting actin automaton 
with a random configuration of `0' and `1' states, where each state is represented with probability 0.5. We evolved the automata $t$ steps and collect linear configurations into a two-dimensional matrix $M = (m_{ij})$, where $1 \leq i \leq n$ and
$1 \leq j \leq \tau$ and $m_{ij}$ is a state of automaton $i$ at time step $j$. The integral measures are
\begin{itemize}
\item Shannon entropy $H$. Let $W$ be a set of all possible configurations of a 9-cell neighbourhood of entry $(i,j)$ of matrix $M$. We calculate number of non-resting configurations as $\eta = \sum_{ a \in M} \epsilon(a)$, where $\epsilon(a)=0$ if for every resting $a$ all its neighbours are resting, and $\epsilon(a)=1$ otherwise. The Shannon entropy is calculated as $- \sum_{w \in W} (\nu(w)/\eta \cdot ln (\nu(w)/\eta))$, where $\nu(w)$ is a number of times the neighbourhood configuration $w$ is found in matrix $M$.
\item Simpson diversity index $D= 1- \sum_{w \in W} (\nu(w)/\eta)^2$.
\item Space filling $P$ is a ratio of non-resting nodes in actin chains in space time configuration of $n \times \tau$ of entities: $P = \sum_{1 \leq i \leq n, 1 \leq j \leq \tau} m_{ij}$.
\item Morphological richness $R$ is a ratio of different $w$ found in $M$ to the total number of possible $w$.
\item Activity $A = (\sum_{1 \leq i \leq n, 1 \leq t \leq \tau} x^t_i) \cdot (n \cdot \tau)^{-1}$,
\item Incoherence $I = |\sum_{1 \leq i \leq n, 1 \leq t \leq \tau} x^t_i - \sum_{1 \leq i \leq n, 1 \leq t \leq \tau} y^t_i | 
\cdot (n \cdot \tau)^{-1}$\\
\item Compressibility $Z = s^{-1}$, where $s$ is a size of space-configuration $M$, in bytes, compressed by LZ algorithm using Java Zip library.
\end{itemize}

These measures are proven to be successful in characterising behaviour of one- and two-dimensional automaton networks, as demonstrated in our previous 
works~\cite{adamatzky2012patterns,adamatzky2012phenomenology, ninagawa2014classifying,redeker2013expressiveness,adamatzky2012diversity}.

  \begin{figure}[!tbp]
 \centering
\includegraphics[width=0.3\textwidth]{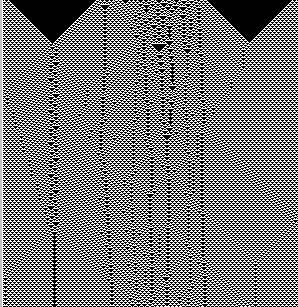}
\caption{Development of Rule $(28,17)$ actin automaton.}
 \label{28_17}
 \end{figure}

The following approach to detect rules supporting localisation was used. We excited resting chains $x$ and $y$ with 
seeds  
 ${\bf x}=(.... 000...0 x_0 x_1 x_2 x_3 x_4 0...000...)$ and 
 ${\bf y}=(.... 000...0 y_0 y_1 y_2 y_3 y_4 0...000...)$ where 
 $x_i, y_j\in  \{ 0, 1 \}$ for $0 \leq i \leq 4$.  For each seed we started the evolution of actin automata from a seed 
 $s=\langle {\bf x}, {\bf y} \rangle$, evolved for $\tau$ steps and calculated activity $A$. 
 If $10 \leq A \leq  6 \cdot \tau$ then $s$  is assumed to generate travelling or stationary localisation. For each rule we counted a number $T$ of seeds generating travelling localizations and a number $S$ of seeds generating stationary localizations.   Only localizations on otherwise resting background were taken into account, as we did not consider localizations travelling in the periodically changing backgrounds, as e.g. one shown in   Fig.~\ref{28_17}. 
 
\begin{figure}[!tbp]
 \centering
\includegraphics[width=0.5\textwidth]{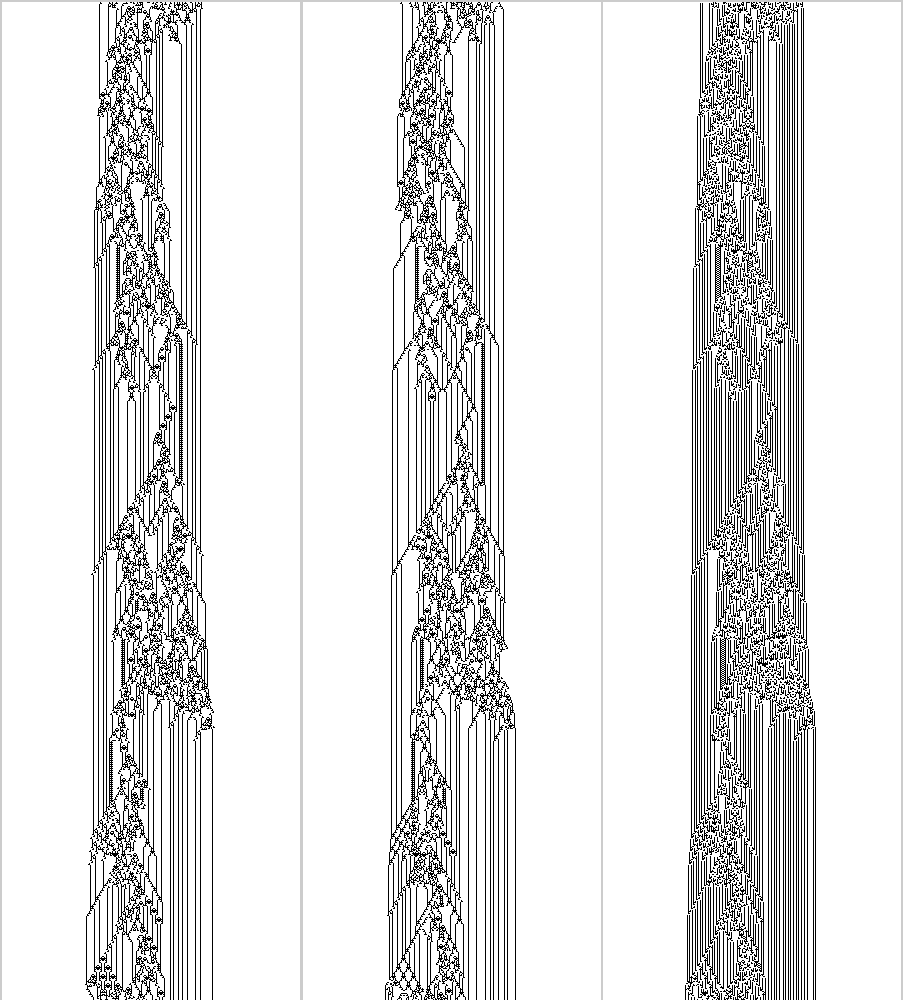}
\caption{Development of Rule $(4,25)$ actin automaton. (Left) Space-time 
configuration of $x$ chain. (Centre) Space-time configuration of $y$ chain.
(Right) Pattern of incoherence between configurations of $x$ and $y$ chains.  }
 \label{4_25}
 \end{figure}
 
 We illustrate the dynamics of actin automata only with space-time configurations of $x$ chains. Morphological differences between configurations of $x$ and $y$ chains do not make a substantial contribution of characterisation of the automaton dynamics (Fig.~\ref{4_25}). Patterns of incoherence indeed give us additional insights into dynamic of actin automata, highlighting a potential topic for further study. 
 
\section{Global characteristics}

  \begin{figure}[!tbp]
 \centering
\subfigure[]{\includegraphics[width=0.6\textwidth]{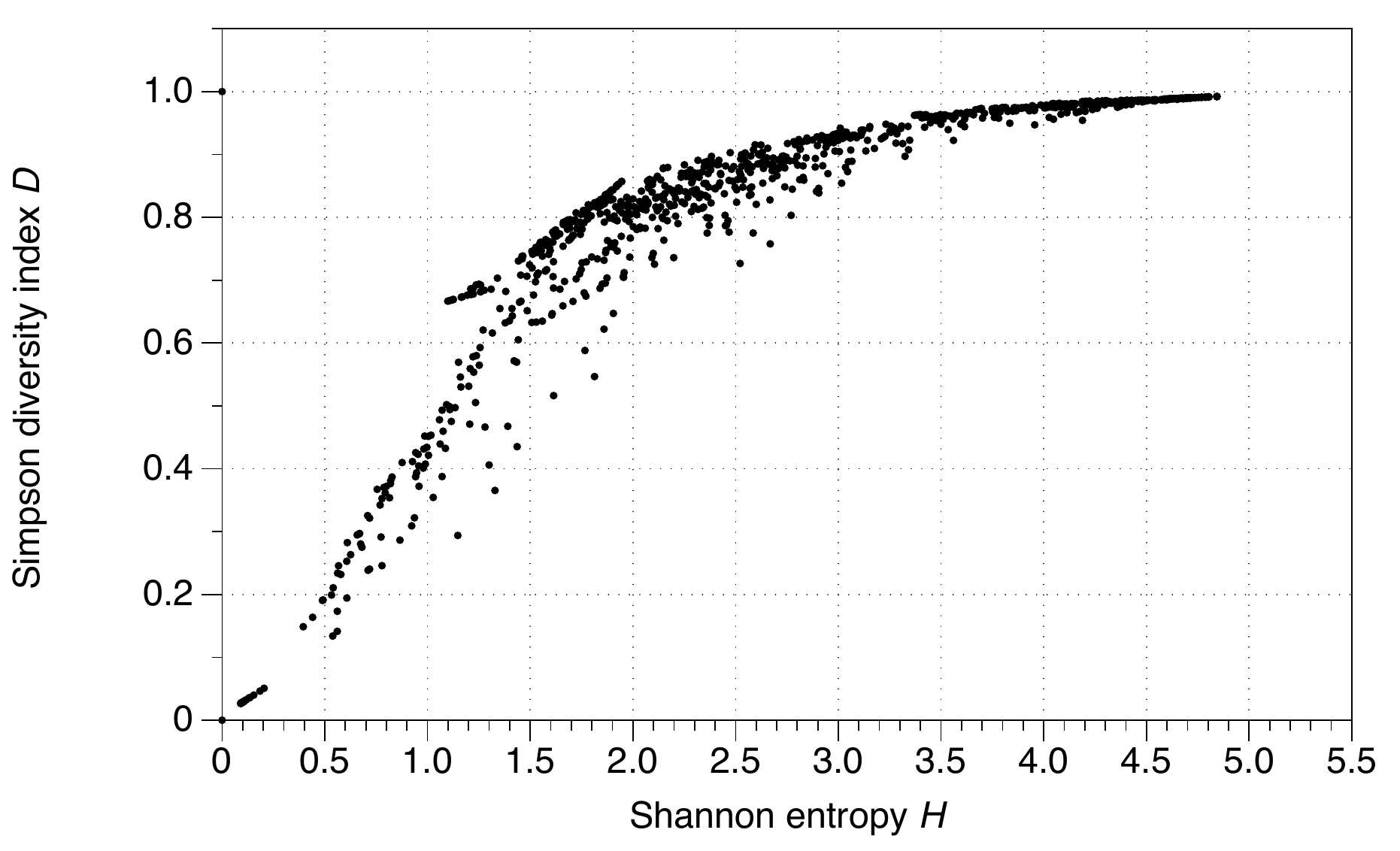}}\\
\subfigure[]{\includegraphics[width=0.33\textwidth]{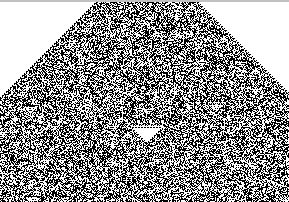}}
\subfigure[]{\includegraphics[width=0.33\textwidth]{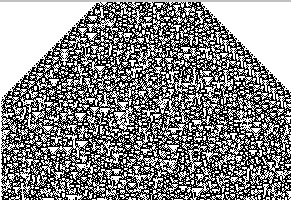}}
\subfigure[]{\includegraphics[width=0.33\textwidth]{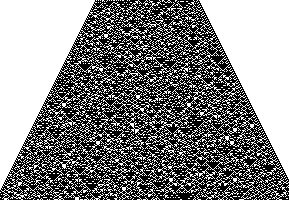}}
\subfigure[]{\includegraphics[width=0.33\textwidth]{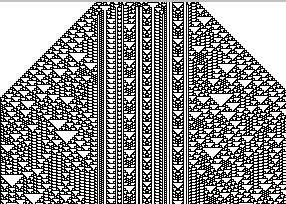}}
\subfigure[]{\includegraphics[width=0.33\textwidth]{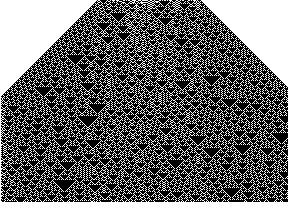}}
\subfigure[]{\includegraphics[width=0.33\textwidth]{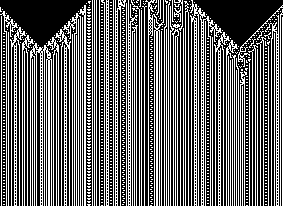}}
\caption{Shannon entropy versus Simpson diversity index. 
(a)~Plot $H$ vs. $D$. 
(b--g)~Space-time configurations developed of automaton $x$ chain governed by exemplar rules: 
(b)~Rule (10,10), $H(10,10)=4.8$,
(c)~Rule (11,6), $H(11,6)=4.5$,
(d)~Rule (7,29), $H(7,29)=4$,
(e)~Rule (11,14), $H(11,14)=3.5$,
(f)~Rule (14,9), $H(14,9)=3$,
(g)~Rule (20,13), $H(20,13)=2.5$.
Time goes down. A node in state `1' is black pixel, and in state `0' is blank. Initially the automata are perturbed by a
random configuration of 100 nodes, where each takes `0' or `1' with probability 0.5. 
}
 \label{entropies}
 \end{figure}
 
   \begin{figure}[!tbp]
 \centering
\includegraphics[width=0.6\textwidth]{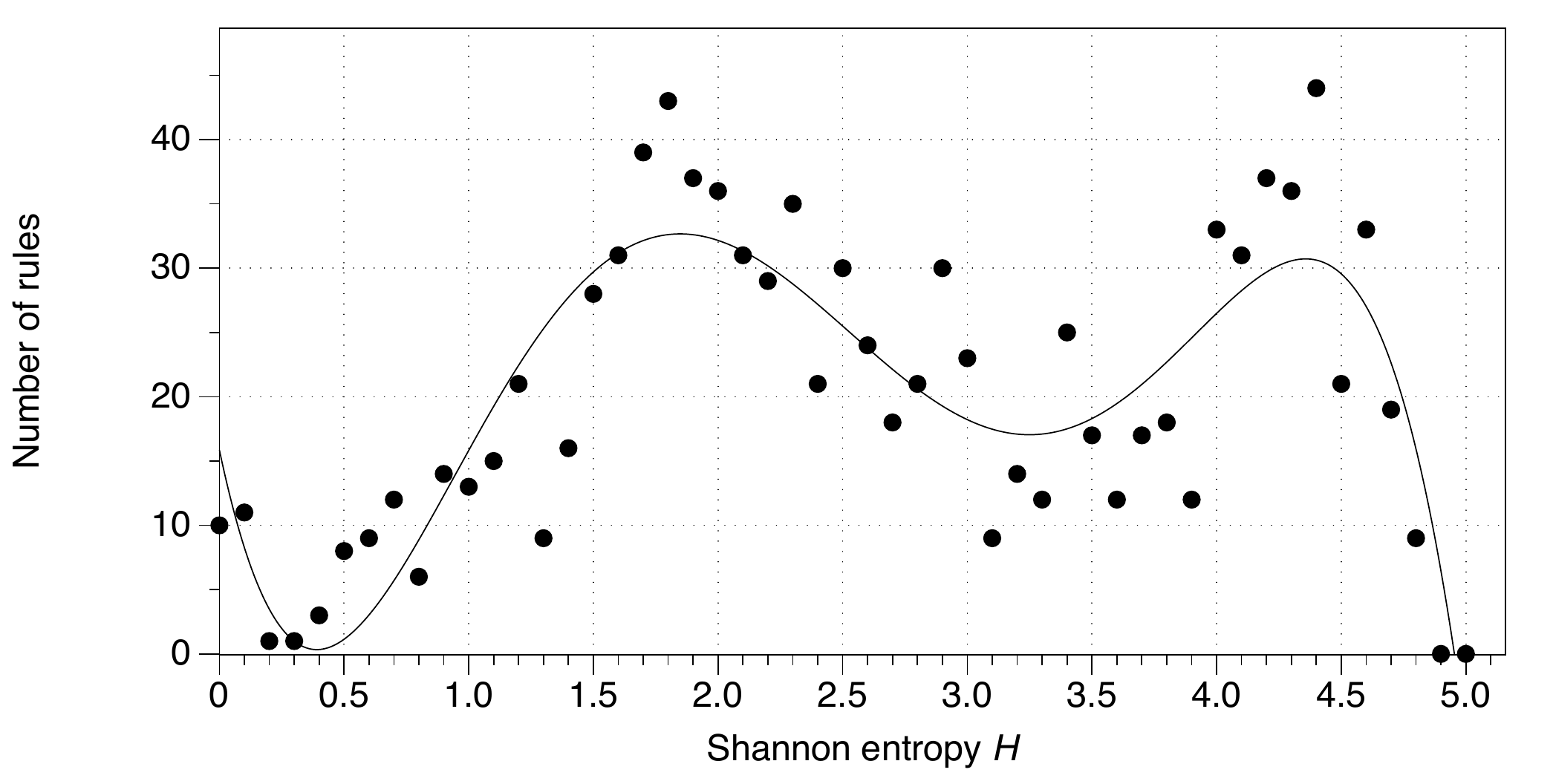}
\caption{Distribution of a number of rules on Shannon entropy $H$ values of the space-time configurations.
The line is an approximation of the distribution by a polynomial of  degree five.}
 \label{entropydistribution}
 \end{figure}


A plot of Shannon entropy $H$ versus Simpson diversity index $D$ is shown in Fig.~\ref{entropies}a.
An accurate  approximation would be polynomial 
$D=0.05+ 0.08*H + 0.62*H^2 - 0.38)*H^3 + 0.08*H^4 -0.01*H^5$ 
although a simpler logarithmic approximation $D =  0.3 \ln(H) + 0.6$ provides a good result with coefficient of determination $R^2=0.905$. Values of $D$ are further provided in this paper for completeness, 
but our statements will consequently be based on $H$. The distribution of rules verses $H$ values is shown in Fig.~\ref{entropydistribution}. The distribution is well approximated by a polynomial with maxima at entropy values around 
1.8 and 4.3. If we arrange rules in ascending order of their values $H$ we find that for $H<1.8$ rules generate 
solid configurations, filled with either `0' or `1' states, or patterns of still localizations in otherwise resting chains. When 
$H$ exceeds 1.8 space-time the chains become filled with regular patterns of activity. Morphology of the space-time become less regular and more complex when $H$ exceeds 4.5. 

Rules with the highest $H$ values exhibit disordered, quasi-chaotic dynamics with no apparently visible structure, e.g. the space-time configuration generated by rule $(10,10)$ is shown in Fig.~\ref{entropies}b--g. The emergence of local domains of regularly arranged states is manifested in decrease of $H$, e.g. the configuration transitions seen in Fig.~\ref{entropies}c to Fig.~\ref{entropies}d: note how the triangles of solid `1'-state domains get their sharp boundaries. Further decreases in entropy are due to the formation of stationary domains of activity co-existing with propagating wave-like patterns (Fig.~\ref{entropies}ef). Stationary localizations dominate the automata chains for low values of the Shannon  entropy (Fig.~\ref{entropies}g).

  \begin{figure}[!tbp]
 \centering
\subfigure[]{\includegraphics[width=0.7\textwidth]{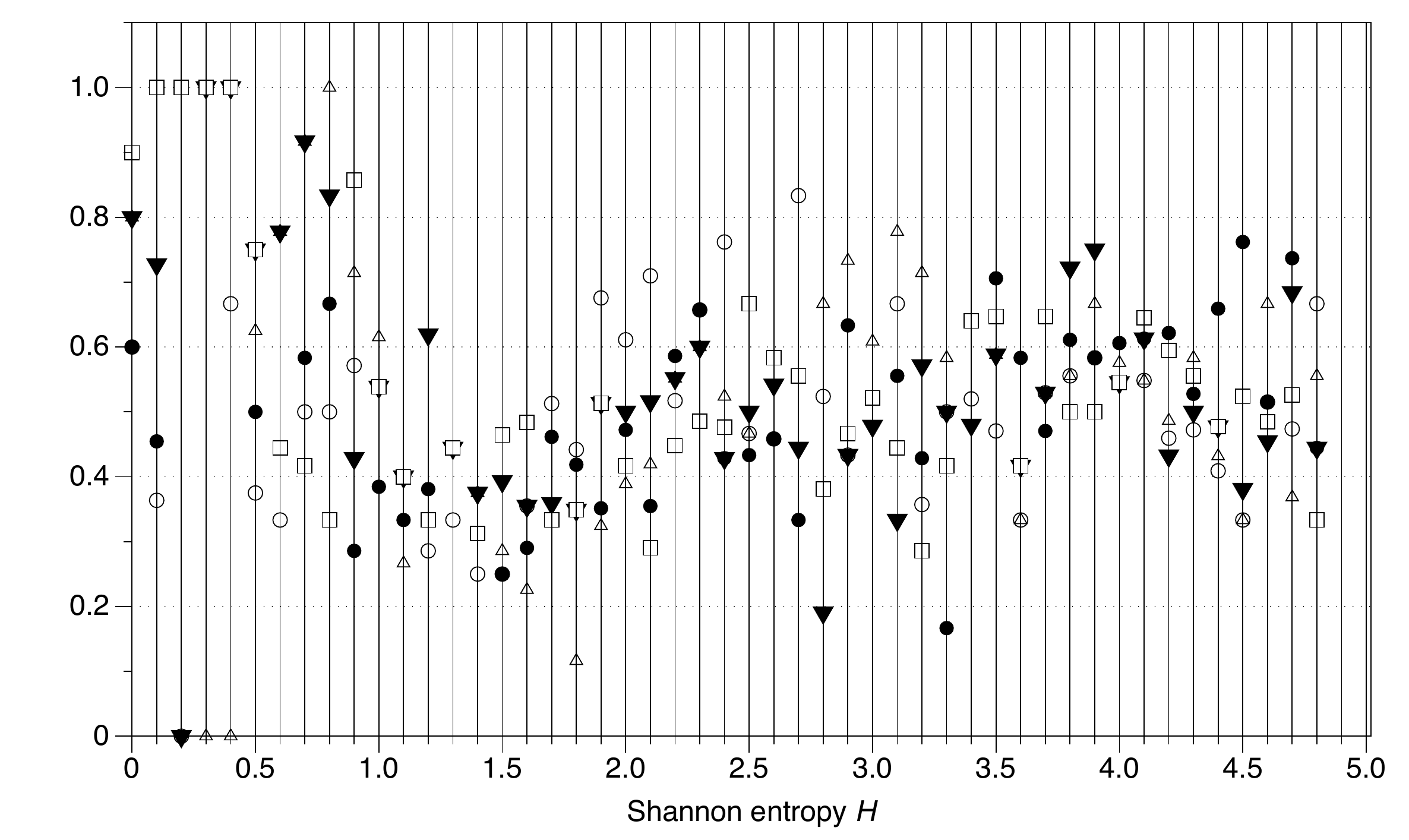}}
\subfigure[]{\includegraphics[width=0.7\textwidth]{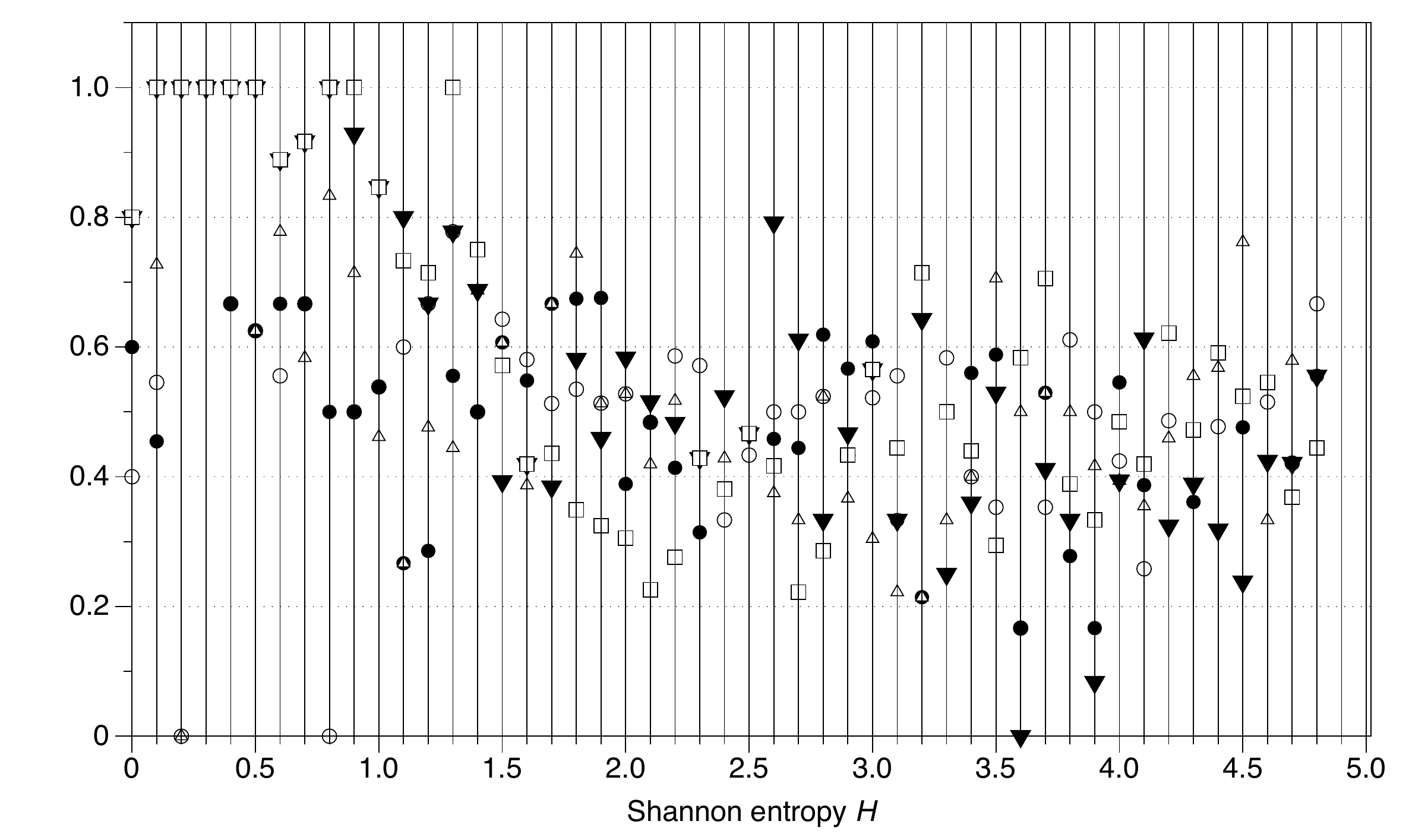}}
\caption{Frequency vectors for rules corresponding Shannon entropy values $H$ ranging from 0 to 1, 
with increment 0.1. (a)~Vector $G_0$ reflecting conditions of node-state transition from state `0' to state `1', 
excitation; and (b)~Vector $G_1$ reflecting conditions of node-state transition from state `1' to state `0', 
retaining excitation. Frequencies are marked as followed: 
$G_{k0}$: empty circle, 
$G_{k1}$: solid disc, 
$G_{k2}$: triangle up, 
$G_{k3}$: triangle down, 
$G_{k4}$: square, where $k=0$ (a) or $k=1$ (b). 
}
 \label{frequencyvectorentropy}
 \end{figure}

Node state transition rules are described by vectors  $F_0$ and $F_1$, where $F_{0i}=1$ means that a resting cell 
becomes excited if it has $i$ excited neighbours. Is there any particular value of $i$, $0 \leq i \leq 4$, which is responsible for space-time configurations having particular value of the Shannon entropy? To answer the question we rank all rules on values of entropies of configurations generated by the functions and split rules in classes 
$L=\{L^1, \cdots, L^50 \}$ of entropy values starting from 0, incrementing with 0.1 and ending at 5. There are between 7 to 30 rules in each of 50 classes of entropy.  Then we calculate frequency vectors frequency vectors $G^z_0$ and $G^z_1$ as follows:
$G^z_{ki} = |L^z|^{-1} \cdot \sum \{ F_{0i}: F_0 \in L^z  \}$,  $k=0, 1$, $|L^z|$ is a size of class $L^z$. That is 
 the frequency vector $G^z_k$ shows how often value '1' appears in vectors $F_k$, $k=0, 1$ for classes of entropy, normalised by size of each class. 
 
 The frequency vectors as a function of Shannon entropy are shown in  Fig.~\ref{frequencyvectorentropy}. Let us exemplify the construction.  In  Fig.~\ref{frequencyvectorentropy}a for $H=0.9$ we see labels in the following order: 
 empty square,  empty triangle up, empty circle, solid triangle down and solid disk. This means that rule-vector $F_0$ 
 corresponding to the rule generating configuration with entropy $H=0.9$ more often has 1 in position 4, $F_{04}=1$,
 less often in position 2, less often in position 0, less often in position 3 and less often in position 1. In words, if 
 a rule generates space-time configurations with entropy $H=0.9$  then more like a resting is excited if it has four neighbours, less like two neighbours, less likely zero neighbours, less likely three neighbours and almost never one neighbour.  To detect what number of neighbours exciting a node is most responsible for generating configurations of 
 certain entropy we select, for each class/interval of entropy, only positions which have maximum frequencies in the class/interval. Thus for a transition from state `0' to state `1' the following numbers of excited neighbours are most critical, they are ordered in increase of $H$: 
 444443324234 344000000110 440222222241 1433141211. The corresponding sequence for a transition from state `1' to state `1' is 444444444434 400121330034 331110401244 0003442420. The sequences are intentionally split into four parts each, the parts corresponds to four equal intervals of the entropy $H$. For each interval we calculate dominating position as follows. For $0 \leq H \leq 1.25$  transitions $0 \rightarrow 1$ and $1 \rightarrow 1$ happen more likely if four neighbours are excited. For entropy  in the interval $]1.25, 2.5]$ node excites or remains excited even if no neighbours are excited, autonomous excitation. Autonomous excitations pose little interest in term communication of signals via travelling localizations, therefore we discard dominating position 0 and look at the sub-sequences 344000000110 (transition $0 \rightarrow 1$) and 400121330034 (transition $1 \rightarrow 1$).  We see that dominating positions for transition $0 \rightarrow 1$ are 1 and 4, and for transition $1 \rightarrow 1$ dominating position is 3.  When 
entropy exceeds 2.5  and yet remains below 3.75 a resting cell excites more likely when it has 2 excited 
neighbours and remains excited if it has 1 excited neighbour. For the highest values of entropy, $3.75 < H \leq 5$ we observe dominating number of neighbours 1 for transition $0 \rightarrow 1$ and 0 for $1 \rightarrow 1$.

\begin{proposition}
Shannon entropy of space-time configurations generated by actin automata is proportional to sensitivity of actin chain nodes: the higher is sensitivity the larger values of entropy the configurations have.
\end{proposition}

With increase of entropy dominating number of neighbours necessary to excite a resting node or to keep excited node excited is decreasing as  4 to 1, 4 to 2 to 1 and 4 to 3 to 1 to 0. This is a reflection of increased
 sensitivity of actin nodes.  Decrease of local excitation necessary to keep a node excited ($4 \rightarrow 3 \rightarrow 1 \rightarrow 0$ can be also interpreted as increase in sustainability of excitation, or even as a transition from lateral excitation to lateral inhibition. Namely, for entropy  in the 
 interval $H \in [0, 1.25]$ an excited node stays excited if it has four excited neighbours and $H \in ]1.25, 2.5]$  if it has three excited neighbours. That is excited nodes in actin automata stimulate each other and thus stay excited longer. The situation is  changed from lateral stimulation to rather lateral inhibition when entropy exceeds 2.5: an excited node remains excited if one neighbour,  $H \in ]2.5, 3.75]$, or no neighbours, $H \in ]3.75, 5]$, are excited; this can be interpreted as that an excessive amount of excited neighbours inhibits excitation of a node.

  \begin{figure}[!tbp]
 \centering
\subfigure[]{\includegraphics[width=0.6\textwidth]{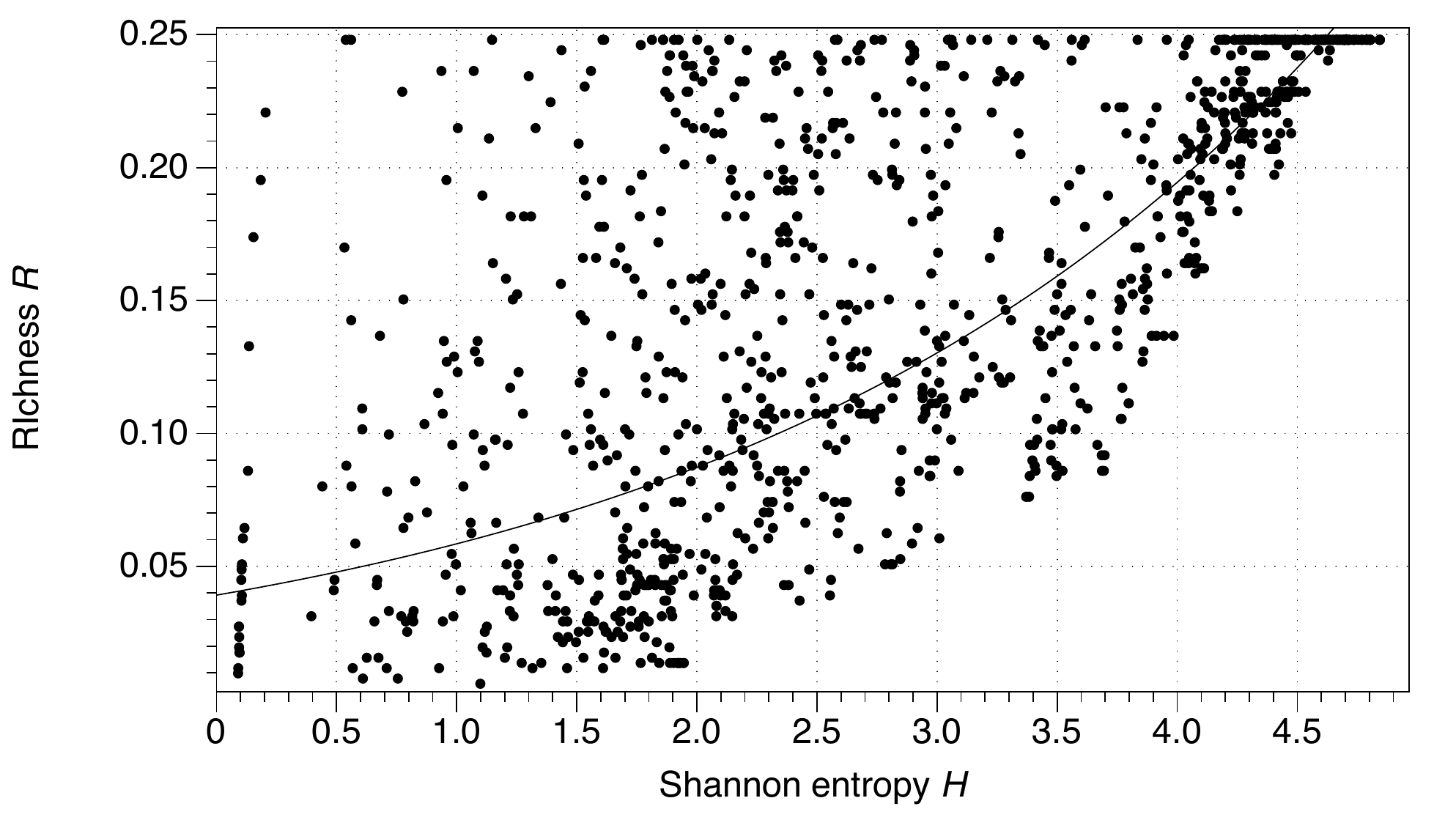}}
\subfigure[]{\includegraphics[width=0.7\textwidth]{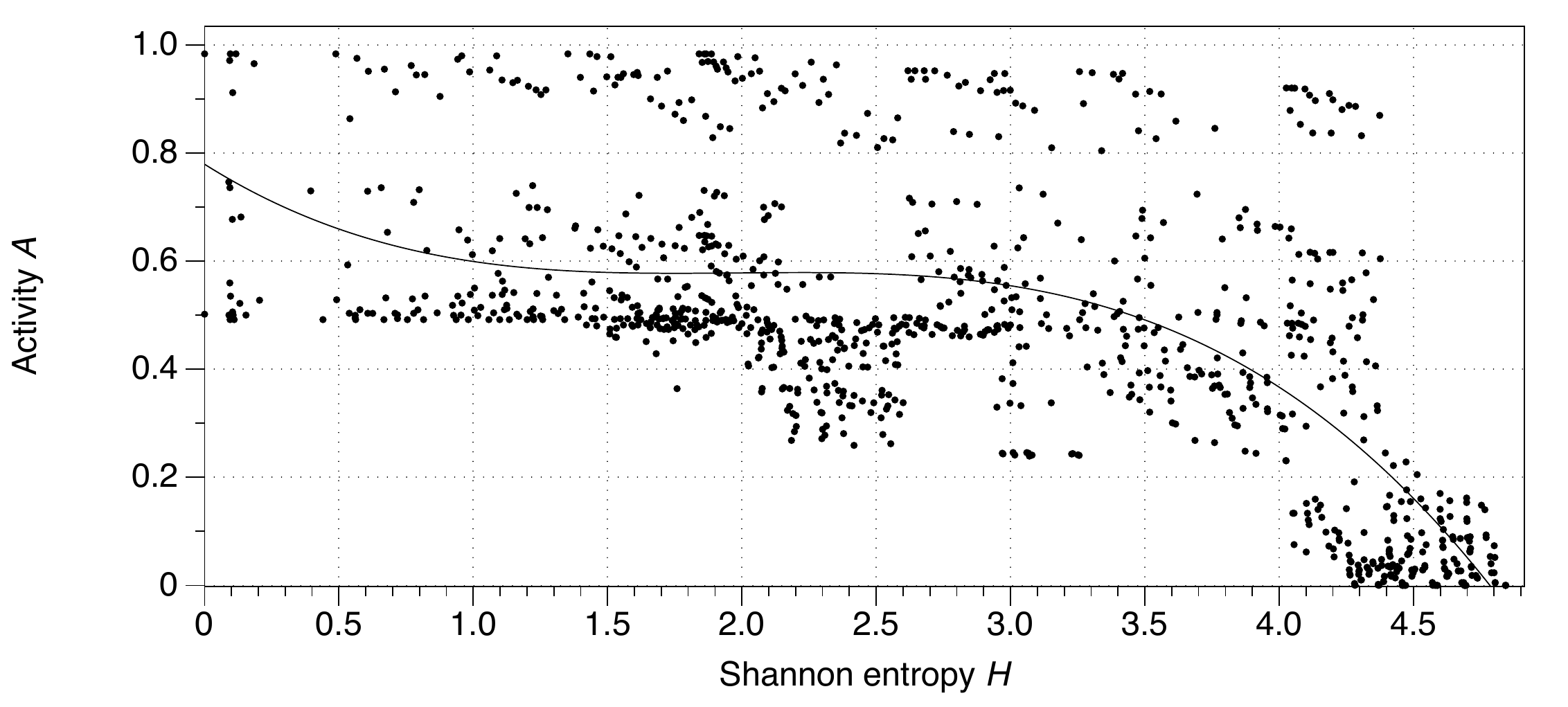}}

\caption{Shannon entropy $H$ versus morphological richness $R$ and activity $A$. 
(a)~$H$ vs. $R$, line is an exponential fit, 
(b)~$H$ vs. $A$, line is a cubic fit, 
}
 \label{shannonrichness}
 \end{figure}

 We could expect that morphological richness $R$, ratios of different $3 \times 3$ patterns in space-time configurations
 will be proportional to entropy $H$. As we can see in Fig.~\ref{shannonrichness}a, this is indeed the case. Richness $R$ grows exponentially with increase of the entropy. Activity $A$ is not a good indicator of morphological complexity the 
 automata dynamics, as we can see in Fig.~\ref{shannonrichness}b, activity remains rather stable for the
 entropy below 3.5 and only starts to substantially drop down when $H$ exceed 4. This is because $A$ depends a ratio of 
 excited nodes and truly complex structures have rather low level of excitations yet elaborately arranged interacting patterns of the excitation.

  \begin{figure}[!tbp]
 \centering
\subfigure[]{\includegraphics[width=0.6\textwidth]{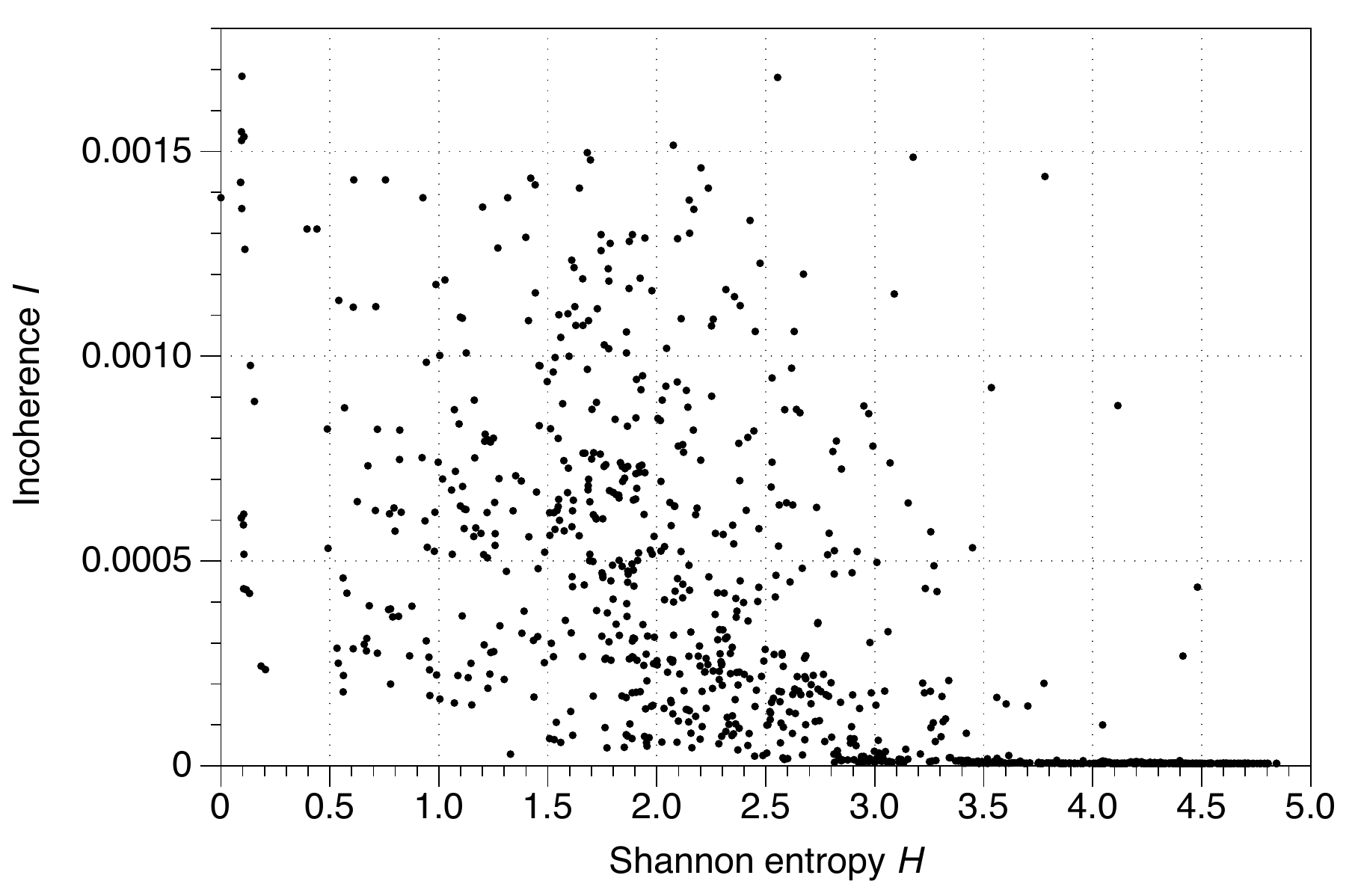}}
\subfigure[]{\includegraphics[width=0.6\textwidth]{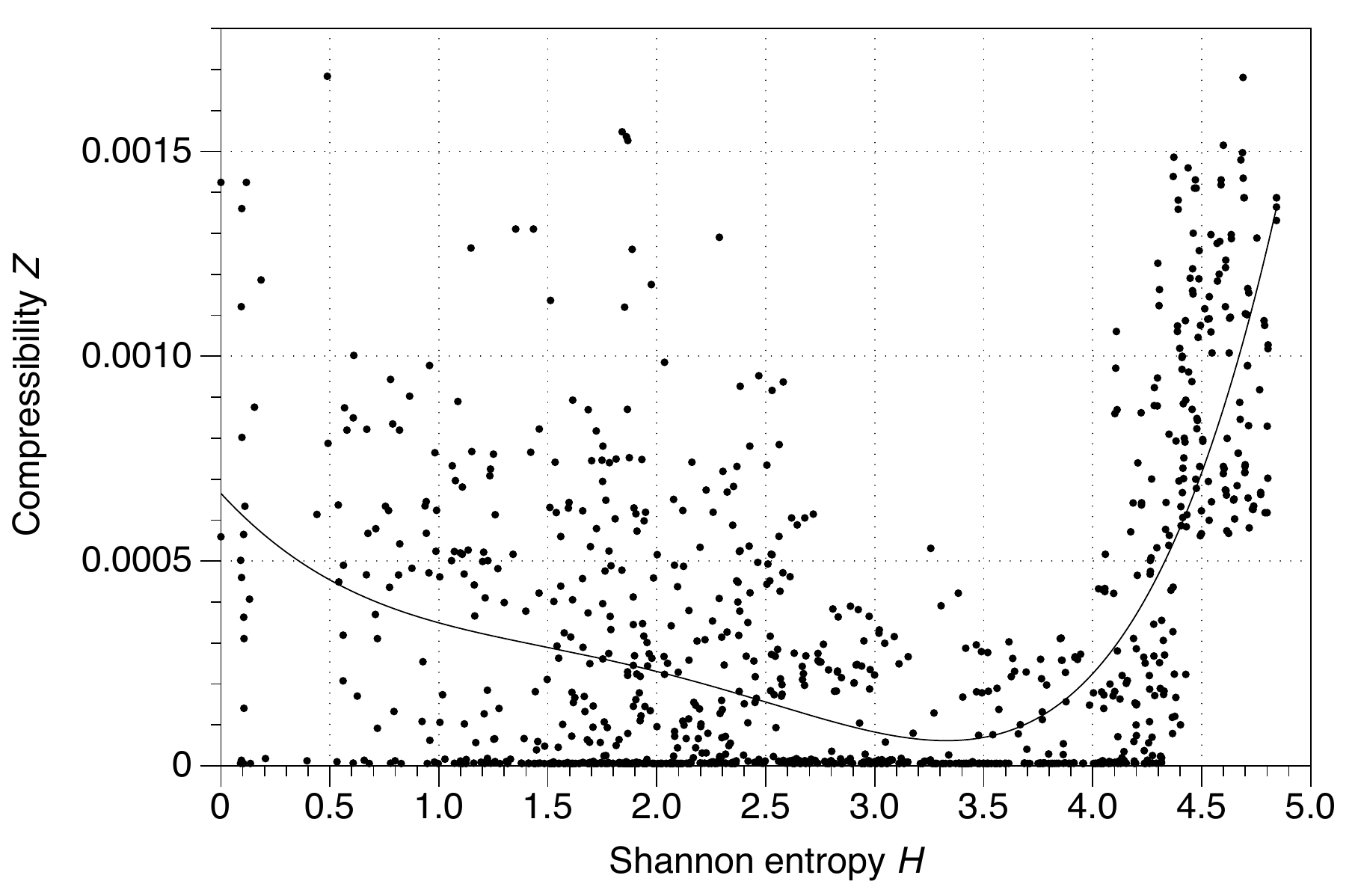}}
\caption{The entropy versus incoherence and compressibility. 
(a)~$H$ vs. $I$, 
.(b)~$H$ vs. $Z$. 
}
 \label{shannonincoherence}
 \end{figure}
 
 There is  a poor correlation between Shannon entropy $H$ and incoherence $I$ (Fig.~\ref{shannonincoherence}a).
 The incoherence indicates how strongly are two chains of actin polymers are desynchronised and might reflect that 
 there are different types of patterns propagating almost independently on the parallel chains of actin units. With regards 
 to compressibility $Z$, the relation could be approximated by quadratic polynomial, where rules with $H$ between 2 and 2 has lowest indicators of compressibility  (Fig.~\ref{shannonincoherence}b).

 \section{Localizations}

 \begin{figure}[!tbp]
 \centering
\includegraphics[width=.6\textwidth]{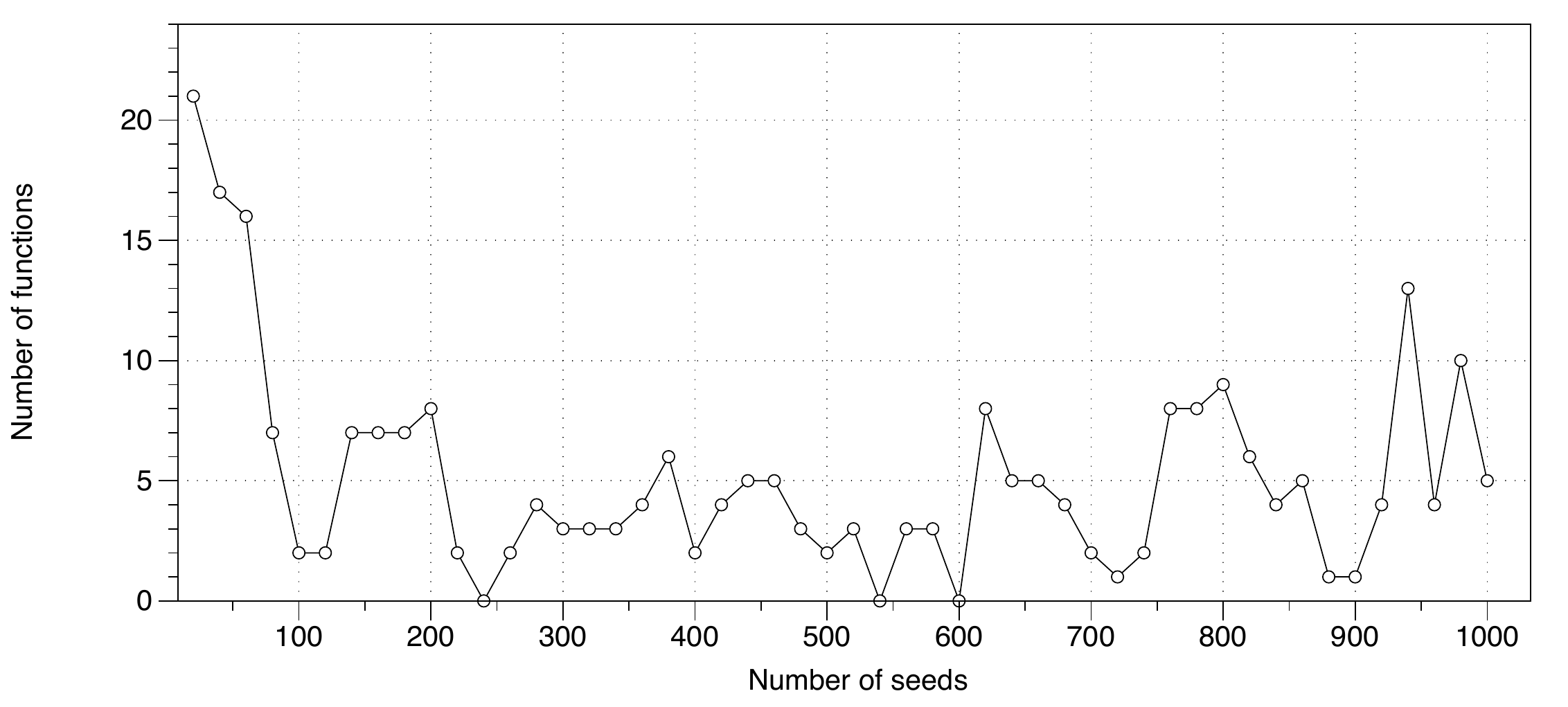}
\caption{Distribution of a number of rules supporting localizations on a number of seeds which leads to the localizations.}
 \label{distributionseeds}
 \end{figure}
 
 Distribution of rules on a number of seeds supporting localizations (Fig.~\ref{distributionseeds}) is quite uniform, there are rules which develop fifty seeds into localizations and there are rules where almost every seeds leads to a stationary or propagating localisation. 705 of 1024 rules, e.g. almost 69\%, do not support any localizations, neither stationary nor travelling. By arranging rules in the descending order on a number of seeds they develop into localisation we can consider three important cases: rules that support traveling localizations, rules that support stationary, rules that support both stationary and mobile localisation. 
 
 \subsection{Travelling localisations}

\begin{table}
\caption{Rules supporting travelling localizations.}
\begin{tiny}
\begin{tabular}{l|llllllll}
	Rule				&	$T$	&	$H$	&	$D$	&	$R$	&	$P$	&	$A$	&	$I$	&	$Z$		\\ \hline
$(	7	,	20	)$	&	271	&	4.31	&	0.98	&	0.22	&	0.84	&	0.32	&	0.28	&	6.67E-006		\\
$(	5	,	26	)$	&	240	&	4.70	&	0.99	&	0.25	&	0.92	&	0.39	&	0.39	&	6.40E-006		\\
$(	4	,	26	)$	&	210	&	4.54	&	0.99	&	0.25	&	0.86	&	0.31	&	0.34	&	6.92E-006		\\
$(	5	,	25	)$	&	197	&	3.83	&	0.95	&	0.25	&	0.73	&	0.14	&	0.19	&	1.37E-005		\\
$(	6	,	20	)$	&	197	&	4.37	&	0.98	&	0.25	&	0.86	&	0.37	&	0.28	&	6.33E-006		\\
$(	8	,	0	)$	&	191	&	3.50	&	0.96	&	0.08	&	0.83	&	0.23	&	0.26	&	6.82E-006		\\
$(	7	,	21	)$	&	179	&	4.31	&	0.98	&	0.22	&	0.97	&	0.56	&	0.39	&	5.97E-006		\\
$(	8	,	1	)$	&	170	&	3.50	&	0.96	&	0.09	&	0.84	&	0.23	&	0.26	&	6.81E-006		\\
$(	8	,	24	)$	&	146	&	3.71	&	0.97	&	0.19	&	0.80	&	0.25	&	0.18	&	7.39E-006		\\
$(	7	,	4	)$	&	145	&	2.75	&	0.92	&	0.20	&	0.54	&	0.06	&	0.07	&	1.82E-004		\\
$(	8	,	16	)$	&	145	&	3.54	&	0.96	&	0.13	&	0.77	&	0.24	&	0.17	&	7.61E-006		\\
$(	7	,	5	)$	&	116	&	4.31	&	0.98	&	0.21	&	0.93	&	0.29	&	0.24	&	8.71E-006		\\
$(	6	,	21	)$	&	110	&	4.40	&	0.98	&	0.25	&	0.70	&	0.14	&	0.14	&	1.28E-005		\\
$(	9	,	0	)$	&	101	&	3.69	&	0.97	&	0.09	&	0.92	&	0.30	&	0.36	&	6.17E-006		\\
$(	9	,	1	)$	&	85	&	3.70	&	0.97	&	0.09	&	0.91	&	0.30	&	0.36	&	6.15E-006		\\
$(	9	,	16	)$	&	85	&	3.63	&	0.96	&	0.14	&	0.81	&	0.26	&	0.18	&	7.18E-006		\\
$(	10	,	16	)$	&	78	&	3.96	&	0.98	&	0.16	&	0.90	&	0.32	&	0.29	&	6.19E-006		\\
$(	10	,	24	)$	&	77	&	4.20	&	0.98	&	0.23	&	0.92	&	0.34	&	0.29	&	6.14E-006		\\
$(	10	,	0	)$	&	72	&	3.69	&	0.97	&	0.09	&	0.93	&	0.30	&	0.39	&	6.15E-006		\\
$(	8	,	2	)$	&	64	&	4.02	&	0.98	&	0.18	&	0.90	&	0.29	&	0.32	&	6.19E-006		\\
$(	8	,	3	)$	&	63	&	4.03	&	0.98	&	0.18	&	0.90	&	0.29	&	0.32	&	6.18E-006		\\
$(	10	,	1	)$	&	58	&	3.69	&	0.97	&	0.09	&	0.92	&	0.30	&	0.40	&	6.14E-006		\\
$(	9	,	24	)$	&	57	&	3.79	&	0.97	&	0.21	&	0.82	&	0.27	&	0.20	&	7.21E-006		\\
$(	5	,	6	)$	&	52	&	4.20	&	0.98	&	0.25	&	0.88	&	0.46	&	0.20	&	6.43E-006		\\
$(	6	,	26	)$	&	44	&	4.67	&	0.99	&	0.25	&	0.96	&	0.49	&	0.45	&	6.19E-006		\\
$(	5	,	24	)$	&	38	&	2.57	&	0.83	&	0.22	&	0.54	&	0.08	&	0.14	&	5.61E-005		\\
$(	8	,	17	)$	&	38	&	4.25	&	0.98	&	0.18	&	0.91	&	0.33	&	0.26	&	6.17E-006		\\
$(	4	,	28	)$	&	37	&	2.83	&	0.86	&	0.19	&	0.94	&	0.50	&	0.39	&	2.35E-005		\\
$(	5	,	28	)$	&	36	&	4.24	&	0.98	&	0.23	&	0.91	&	0.41	&	0.34	&	6.11E-006		\\
$(	8	,	8	)$	&	36	&	4.26	&	0.98	&	0.20	&	0.93	&	0.32	&	0.39	&	6.26E-006		\\
$(	5	,	21	)$	&	32	&	3.31	&	0.92	&	0.25	&	0.36	&	0.06	&	0.08	&	1.08E-005		\\
$(	8	,	9	)$	&	32	&	4.31	&	0.98	&	0.21	&	0.93	&	0.33	&	0.40	&	6.17E-006		\\
$(	9	,	2	)$	&	32	&	4.11	&	0.98	&	0.19	&	0.94	&	0.31	&	0.36	&	6.05E-006		\\
$(	9	,	3	)$	&	32	&	4.13	&	0.98	&	0.19	&	0.94	&	0.32	&	0.36	&	6.02E-006		\\
$(	10	,	17	)$	&	31	&	4.38	&	0.99	&	0.21	&	0.95	&	0.37	&	0.36	&	5.91E-006		\\
$(	10	,	25	)$	&	29	&	4.57	&	0.99	&	0.25	&	0.94	&	0.39	&	0.35	&	5.90E-006		\\
$(	6	,	10	)$	&	28	&	4.62	&	0.99	&	0.25	&	0.90	&	0.46	&	0.45	&	6.36E-006		\\
$(	8	,	25	)$	&	27	&	4.28	&	0.98	&	0.22	&	0.91	&	0.33	&	0.28	&	6.25E-006		\\
$(	10	,	2	)$	&	27	&	4.19	&	0.98	&	0.23	&	0.95	&	0.34	&	0.42	&	5.92E-006		\\
$(	10	,	3	)$	&	27	&	4.28	&	0.98	&	0.22	&	0.95	&	0.36	&	0.41	&	5.90E-006		\\
$(	6	,	24	)$	&	25	&	3.35	&	0.92	&	0.21	&	0.84	&	0.16	&	0.25	&	2.00E-005		\\
$(	4	,	27	)$	&	22	&	4.20	&	0.97	&	0.25	&	0.88	&	0.27	&	0.23	&	1.02E-005		\\
$(	6	,	11	)$	&	22	&	1.07	&	0.39	&	0.24	&	0.98	&	0.84	&	0.04	&	1.54E-004		\\
$(	7	,	24	)$	&	22	&	3.96	&	0.97	&	0.19	&	0.93	&	0.43	&	0.71	&	1.27E-005		\\
$(	4	,	6	)$	&	21	&	2.18	&	0.85	&	0.23	&	0.97	&	0.60	&	0.58	&	1.21E-004		\\
$(	6	,	5	)$	&	20	&	3.60	&	0.95	&	0.25	&	0.47	&	0.07	&	0.08	&	1.51E-004		\\
$(	10	,	8	)$	&	20	&	4.48	&	0.99	&	0.23	&	0.95	&	0.37	&	0.39	&	5.90E-006		\\
$(	11	,	16	)$	&	19	&	4.03	&	0.98	&	0.16	&	0.95	&	0.36	&	0.34	&	5.98E-006		\\
$(	10	,	9	)$	&	18	&	4.61	&	0.99	&	0.25	&	0.96	&	0.39	&	0.41	&	5.84E-006		\\
$(	11	,	24	)$	&	17	&	4.27	&	0.98	&	0.23	&	0.96	&	0.40	&	0.35	&	6.00E-006		\\
$(	6	,	4	)$	&	16	&	2.55	&	0.90	&	0.23	&	0.44	&	0.05	&	0.04	&	4.65E-004		\\
$(	6	,	25	)$	&	15	&	3.25	&	0.94	&	0.23	&	0.98	&	0.61	&	0.33	&	1.03E-005		\\
$(	8	,	21	)$	&	15	&	4.53	&	0.99	&	0.23	&	0.95	&	0.39	&	0.41	&	5.87E-006		\\
$(	4	,	21	)$	&	13	&	3.05	&	0.87	&	0.25	&	0.39	&	0.03	&	0.06	&	1.83E-004		\\
$(	5	,	20	)$	&	13	&	2.74	&	0.85	&	0.25	&	0.39	&	0.03	&	0.06	&	1.87E-004		\\
$(	5	,	11	)$	&	11	&	1.81	&	0.55	&	0.25	&	0.45	&	0.19	&	0.04	&	3.46E-004		\\
$(	7	,	10	)$	&	11	&	4.35	&	0.98	&	0.22	&	0.94	&	0.50	&	0.46	&	6.47E-006		\\
$(	4	,	25	)$	&	10	&	3.34	&	0.91	&	0.23	&	0.60	&	0.10	&	0.14	&	1.98E-005		\\
$(	9	,	21	)$	&	10	&	4.57	&	0.99	&	0.25	&	0.96	&	0.40	&	0.42	&	5.83E-006		\\
$(	4	,	20	)$	&	8	&	2.57	&	0.84	&	0.25	&	0.38	&	0.03	&	0.06	&	2.71E-006		\\
$(	7	,	8	)$	&	8	&	1.84	&	0.69	&	0.17	&	0.92	&	0.42	&	0.83	&	1.71E-006		\\
$(	5	,	10	)$	&	6	&	4.41	&	0.98	&	0.25	&	0.09	&	0.00	&	0.00	&	2.68E-004		\\
$(	8	,	10	)$	&	6	&	4.48	&	0.98	&	0.23	&	0.96	&	0.37	&	0.45	&	5.92E-006		\\
$(	8	,	11	)$	&	6	&	4.43	&	0.98	&	0.23	&	0.96	&	0.39	&	0.44	&	5.97E-006		\\
$(	5	,	27	)$	&	5	&	0.54	&	0.13	&	0.25	&	0.98	&	0.32	&	0.04	&	2.51E-004		\\
$(	4	,	22	)$	&	4	&	2.62	&	0.89	&	0.24	&	0.93	&	0.41	&	0.39	&	1.74E-004		\\
$(	5	,	4	)$	&	4	&	2.95	&	0.92	&	0.23	&	0.25	&	0.02	&	0.02	&	8.79E-004		\\
$(	5	,	5	)$	&	4	&	3.45	&	0.95	&	0.25	&	0.24	&	0.04	&	0.04	&	5.32E-004		\\
$(	5	,	14	)$	&	3	&	2.80	&	0.88	&	0.12	&	0.96	&	0.37	&	0.25	&	7.01E-005		\\
$(	4	,	10	)$	&	2	&	4.48	&	0.98	&	0.25	&	0.05	&	0.00	&	0.00	&	4.36E-004		\\
$(	7	,	9	)$	&	2	&	4.19	&	0.98	&	0.22	&	0.95	&	0.48	&	0.58	&	8.30E-006		\\
$(	7	,	11	)$	&	2	&	0.78	&	0.25	&	0.15	&	0.98	&	0.85	&	0.01	&	1.99E-006		\\
$(	7	,	19	)$	&	1	&	2.93	&	0.90	&	0.15	&	0.93	&	0.48	&	0.17	&	1.40E-004		\\
$(	8	,	20	)$	&	1	&	4.42	&	0.98	&	0.23	&	0.95	&	0.37	&	0.37	&	5.99E-006		\\
$(	12	,	0	)$	&	1	&	2.85	&	0.92	&	0.08	&	0.84	&	0.24	&	0.10	&	1.45E-005		\\
$(	12	,	1	)$	&	1	&	2.85	&	0.92	&	0.08	&	0.84	&	0.24	&	0.27	&	1.25E-005		\\
$(	12	,	16	)$	&	1	&	2.85	&	0.92	&	0.09	&	0.84	&	0.24	&	0.13	&	1.45E-005		\\
\end{tabular}
\end{tiny}
\label{tabletravelling}
\end{table}

  \begin{figure}[!tbp]
 \centering
\includegraphics[width=0.6\textwidth]{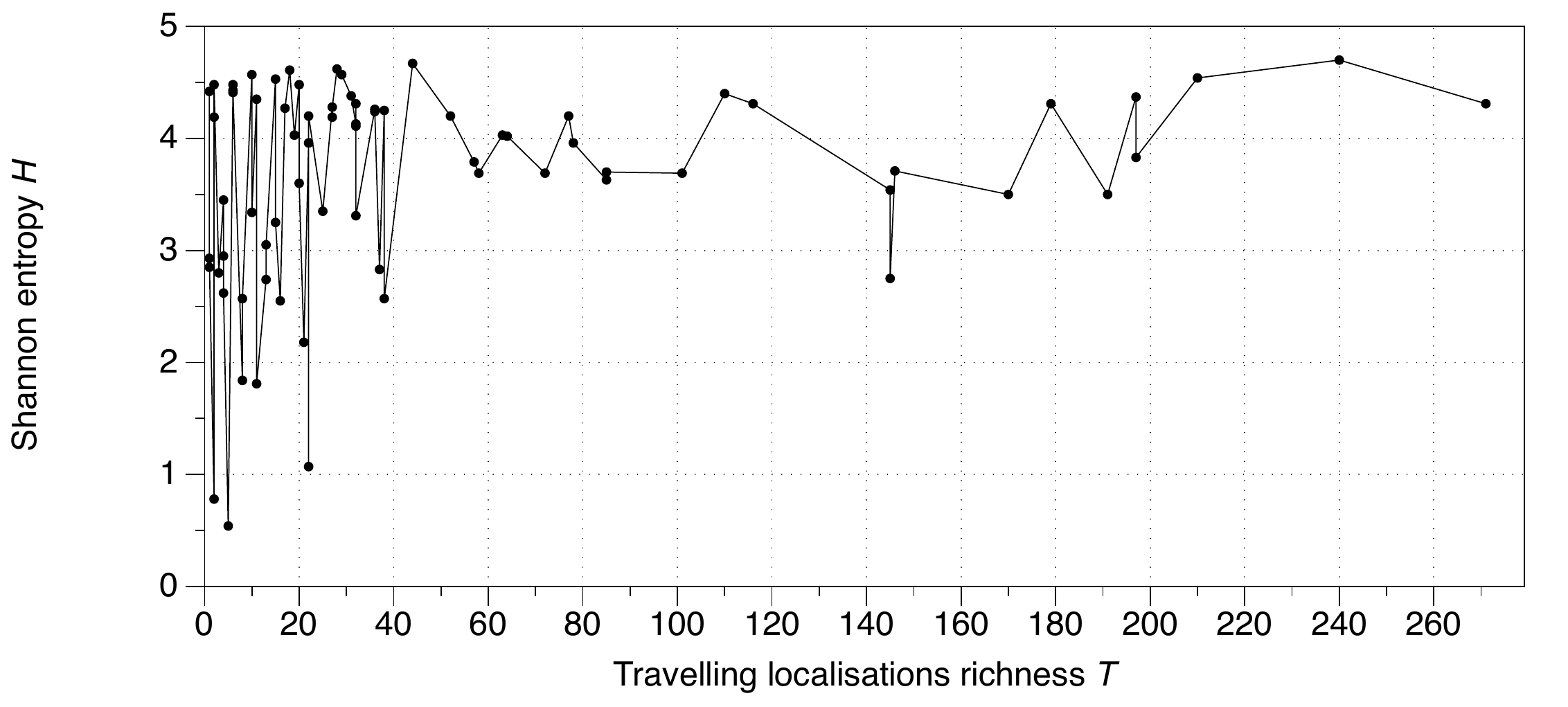}
\caption{Travelling localizations richness $T$ versus Shannon entropy $H$}
 \label{TravellingvsShannon}
 \end{figure}

Rules supporting travelling localizations  and the integral characteristics of the rules 
are shown in Tab.~\ref{tabletravelling}. A plot of entropy $H$ versus $T$ (Fig.~\ref{TravellingvsShannon})
shows that rules with lowest values of the localizations richness $T$  show high dispersion in $H$, they  
exhibit oscillations in entropy, with amplitude 
circa 1.5. The rules with $T$ exceeding 40, i.e. those which develop 40 seeds into the localizatios, show entropy 
values around 4.

\begin{proposition}
Higher values of Shannon entropy are typical for rules supporting large number of travelling localizations.
\end{proposition}

\begin{table}
\caption{Top 14 rules supporting travelling localizations with $T>100$.}
\begin{tabular}{llll}
	Rule				& $F^0$ & $F^1$ & $T$	 \\ \hline
$(	7	,	20	)$	& 00111 & 10100	& 271	\\
$(	5	,	26	)$	& 00101 & 11010			&240	\\
$(	4	,	26	)$	& 00100 & 11010			&210	\\
$(	5	,	25	)$	& 00101 & 11001		&197	\\
$(	6	,	20	)$	& 00110 & 10100			&197	\\
$(	8	,	0	)$	& 01000 & 00000 			&191	\\
$(	7	,	21	)$	& 00111 & 10101			&179	\\
$(	8	,	1	)$	& 01000 & 00001			&170	\\
$(	8	,	24	)$	& 01000 & 11000			&146	\\
$(	7	,	4	)$	& 00111 & 01110			&145	\\
$(	8	,	16	)$	& 01000 & 10000			&145	\\
$(	7	,	5	)$	& 00111 & 00101			&116	\\
$(	6	,	21	)$	& 00110 & 10101			&110	\\
$(	9	,	0	)$	& 01001 & 00000			&101	\\	
\end{tabular}
\label{top15travelling}
\end{table}

Top 14 rules, with values of $T$ exceeding 100 are shown in Tab.~\ref{top15travelling}.

  \begin{figure}[!tbp]
 \centering
\subfigure[]{\includegraphics[width=0.3\textwidth]{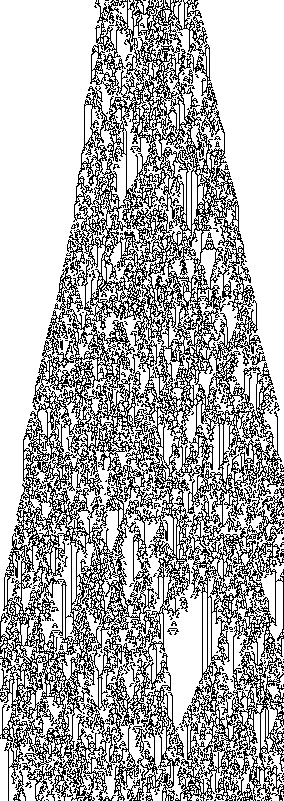}}
\subfigure[]{\includegraphics[width=0.3\textwidth]{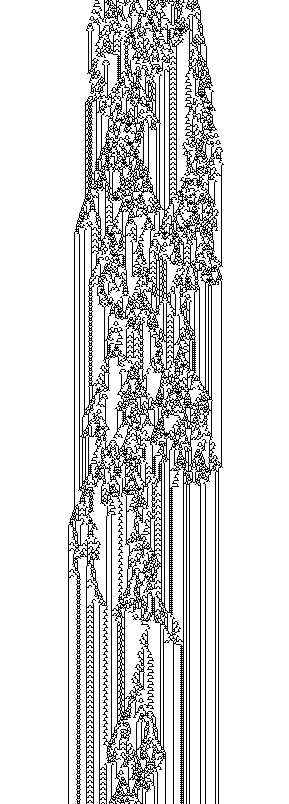}}
\caption{Exemplar space-time configurations generated by (a) Rule (4,26) and (b) Rule (5,25).}
 \label{rule426525}
 \end{figure}

  \begin{figure}[!tbp]
 \centering
\subfigure[]{\includegraphics[width=0.3\textwidth]{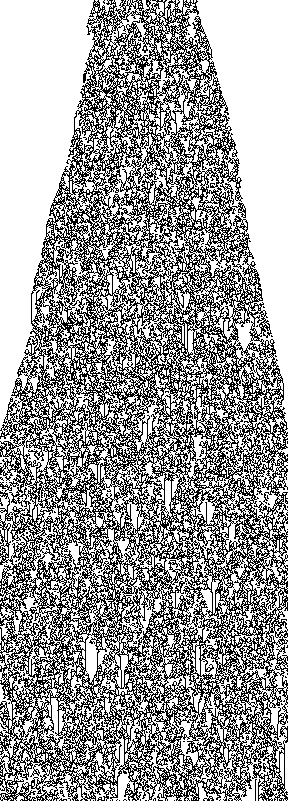}}
\subfigure[]{\includegraphics[width=0.3\textwidth]{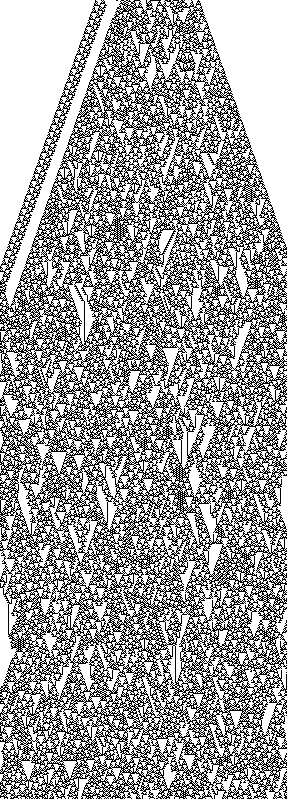}}
\caption{Exemplar space-time configurations generated by (a) Rule (5,26) and (b) Rule (6,20).}
 \label{rule526620}
 \end{figure}

  \begin{figure}[!tbp]
 \centering
\subfigure[]{\includegraphics[width=0.3\textwidth]{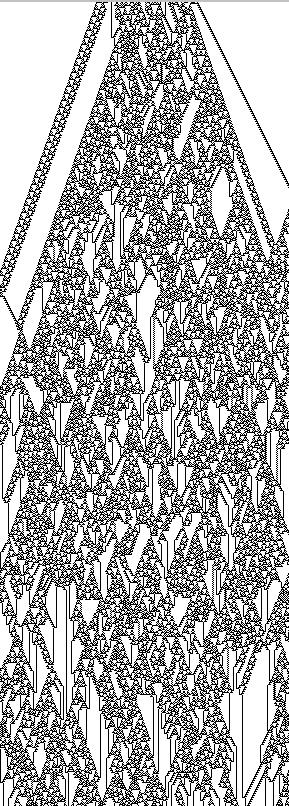}}
\subfigure[]{\includegraphics[width=0.3\textwidth]{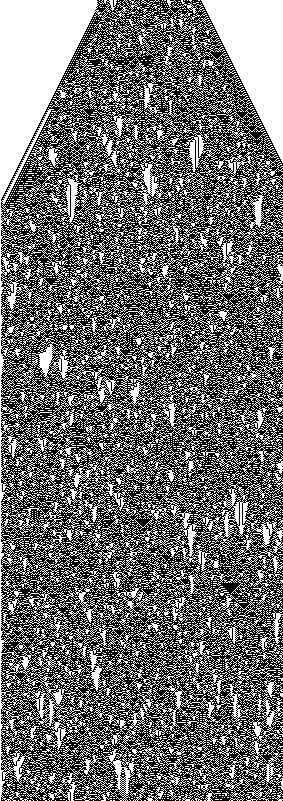}}
\caption{Exemplar space-time configurations generated by (a) Rule (7,20) and (b) Rule (7,21).}
 \label{rule720721}
 \end{figure} 
 

Exemplar space-time configurations generated by some rules from Tab.~\ref{top15travelling} are shown in 
Figs.~\ref{rule426525}, \ref{rule526620} and \ref{rule720721}. There actin automata were excited a random configuration (100 nodes) of state `1' and `0' assigned to nodes with probability 0.5. In most case we see that original pattern of perturbation spreads on the actin chain. The pattern front can propagate with a speed of light (one per node per iteration) as e.g. in Rule (7,20) in Fig.~\ref{rule720721}a, or slightly slower, e.g. Rule (4,26) in Fig~\ref{rule426525}a.
In majority of the rules Tab.~\ref{top15travelling} initial perturbation produces a great variety of stationary and travelling which collide with each, form new localizations, etc. and thus the space-configurations are well filled with excitation activities. In some cases, e.g. Rule (5,25) in Fig.~\ref{rule426525}, the original perturbation does not propagate: colliding travelling localizations produce stationary, still, patterns, which in turn limit further propagation of new travelling localizations. In such rules, travelling localizations bounce between still localizations.

 \begin{figure}[!tbp]
 \centering
\subfigure[]{\includegraphics[width=0.5\textwidth]{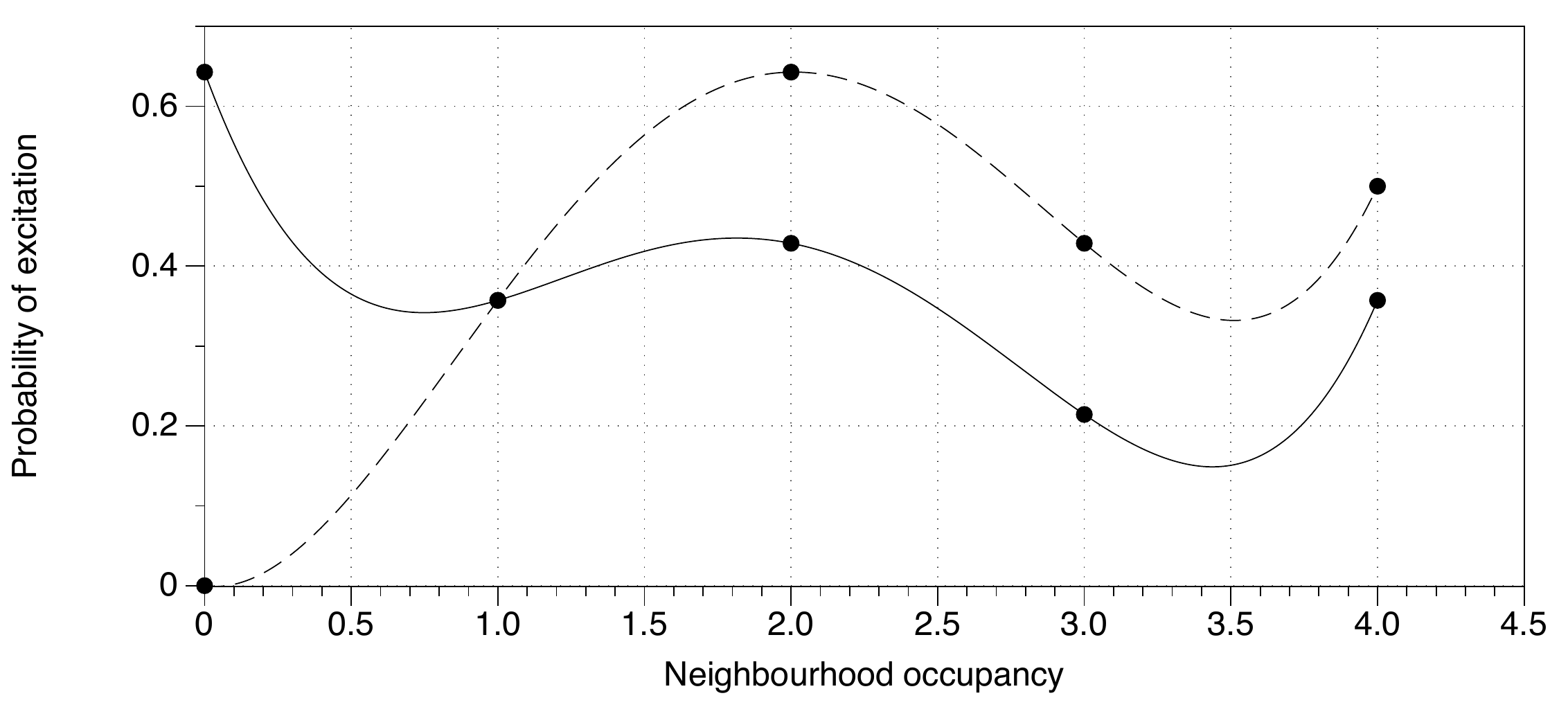}}
 \subfigure[]{\includegraphics[width=0.49\textwidth]{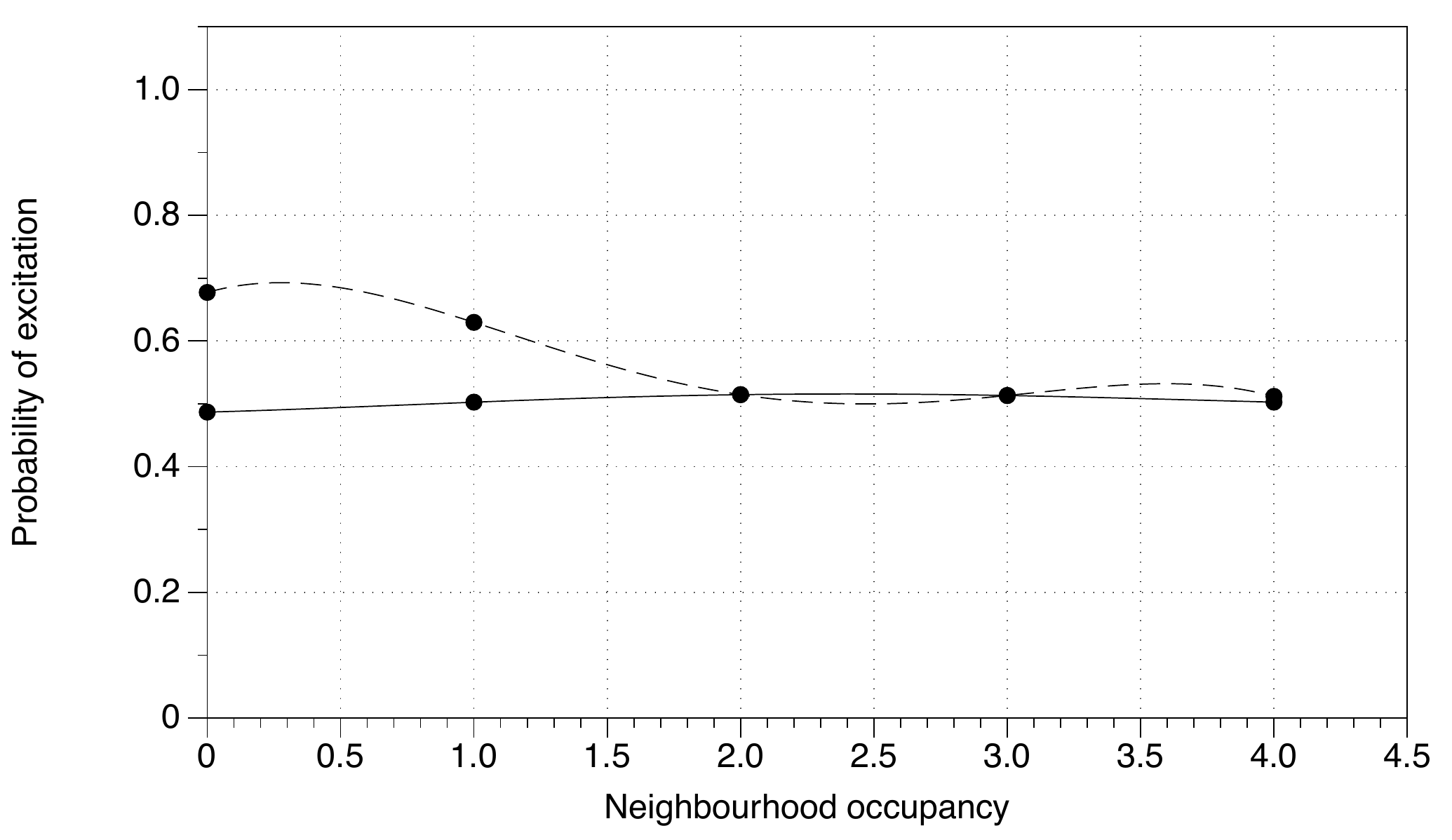}}
  \caption{Dependence of a probability of excitation of a node in actin automaton (a)~supporting travelling 
  localizations and (b)~not supporting any localizations (neither stationary nor travelling) 
  on a number of excited neighbours of the node. The polynomial of the probability (a) is calculated on 14 rules that exhibit travelling localizations for largest number of seeds, and the polynomial of the probability (b) is calculated on 705 rules not supporting any localizations.   Dashed line shows probability of excitation of a resting node, solid line of an excited node, i.e. of an excited node to remain excited.}
 \label{travellingpolynomial}
 \end{figure}

Let us calculate frequency vectors $V^0$ and $V^1$ from $F^0$ and $F^1$. There are fourteen rules in Tab.~\ref{top15travelling},
 $V^0_i$ is a ratio of `1' in $F^0_i$ to 7, the same for $V_1$. 
 Thus we have $V^0 = ( 0 \frac{5}{14} \frac{9}{14} \frac{6}{4} \frac{7}{14})$ and 
 $V^1 = (\frac{9}{14} \frac{5}{14} \frac{6}{14} \frac{3}{14} \frac{5}{14})$. Probabilities of excitation
 can be represented via a polynomial approximation of vectors $V^0$ and $V^1$ shown in Fig.~\ref{travellingpolynomial}a, compare with the probability of excitation of the rules not supporting any localizations 
 in Fig.~\ref{travellingpolynomial}b.
 
  Assuming $0.5$ is a cut off frequency, we can transform the vector to a simplified form: 
 $^*V^0 = (0 0 1 0 1)$ and $^*V^1 = (1 0 0 0 0)$. 
 
 \begin{proposition}
 A rule more likely supports large number of traveling localizations if it has the following node state transitions. A resting node becomes excited if it has two or four excited neighbours.   An excited node remains excited if it has no excited neighbours. 
 \end{proposition}

 \subsection{Stationary localisations}
 
  \begin{figure}[!tbp]
 \centering
\includegraphics[width=0.49\textwidth]{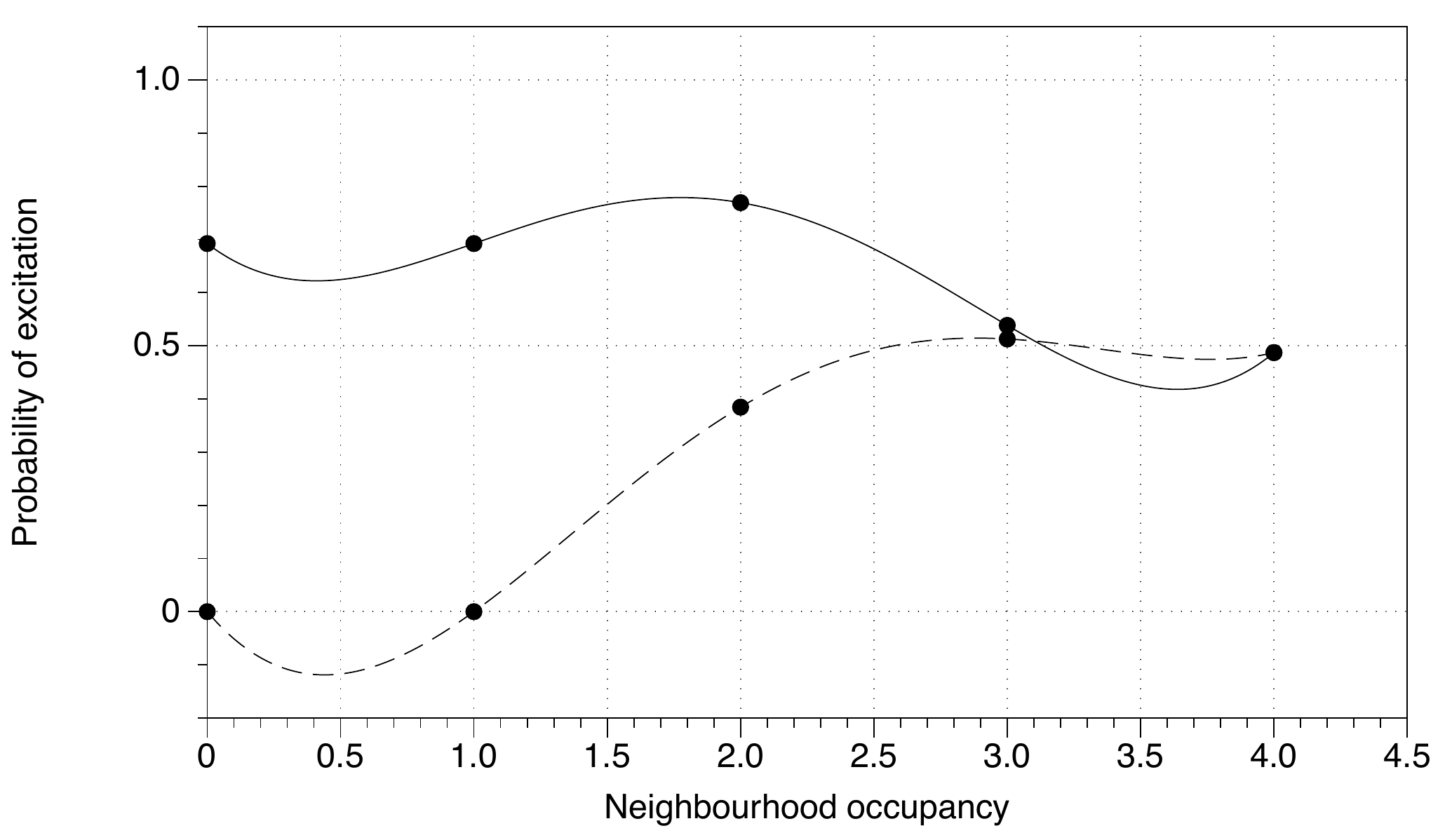}
\caption{Dependence of a probability of excitation of a node in actin automaton supporting stationary 
  localizations  on a number of excited neighbours of the node. The polynomial of the probability is calculated on 39 rules that exhibit stationary localizations for 900 seeds. Dashed line shows probability of excitation of a resting node, solid line of an excited node, i.e. of an excited node to remain excited.}
 \label{stationarypolynomial}
 \end{figure}

Let us calculate frequency vectors $V^0$ and $V^1$ from $F^0$ and $F^1$ for rules supporting stationary 
localizations. There are 39  rules that exhibit stationary localizations for over 900 (of 1024) seeds. 
 $V^0_i$ is a ratio of '1' in $F^0_i$ to 39, the same for $V_1$. 
 Thus we have $V^0 = ( 0  0 \frac{15}{39} \frac{20}{39} \frac{19}{39} )$ and 
 $V^1 = (\frac{27}{39} \frac{27}{39} \frac{30}{39} \frac{21}{39} \frac{19}{39})$. Probabilities of excitation
 can be represented via a polynomial approximation of vectors $V^0$ and $V^1$ shown in Fig.~\ref{stationarypolynomial}a, compare with the probability of excitation of the rules that support travelling 
 localizations in Fig.~\ref{travellingpolynomial}b and rules that do not support any localizations 
 in Fig.~\ref{travellingpolynomial}b. Assuming $0.5$ is a cut off frequency, we can transform the vectors to a simplified form:   $^*V^0 = (0 0 0 1 0)$ and $^*V^1 = (1 1 1 1 0)$.

 \begin{proposition}
 A rule more likely supports large number of stationary localizations  if it has the following node state transitions. A resting node becomes excited only if three of its four  neighbours are excited.  An excited node remains excited if it has less four excited neighbours. 
 \end{proposition}

  \begin{figure}[!tbp]
 \centering
\includegraphics[width=0.5\textwidth]{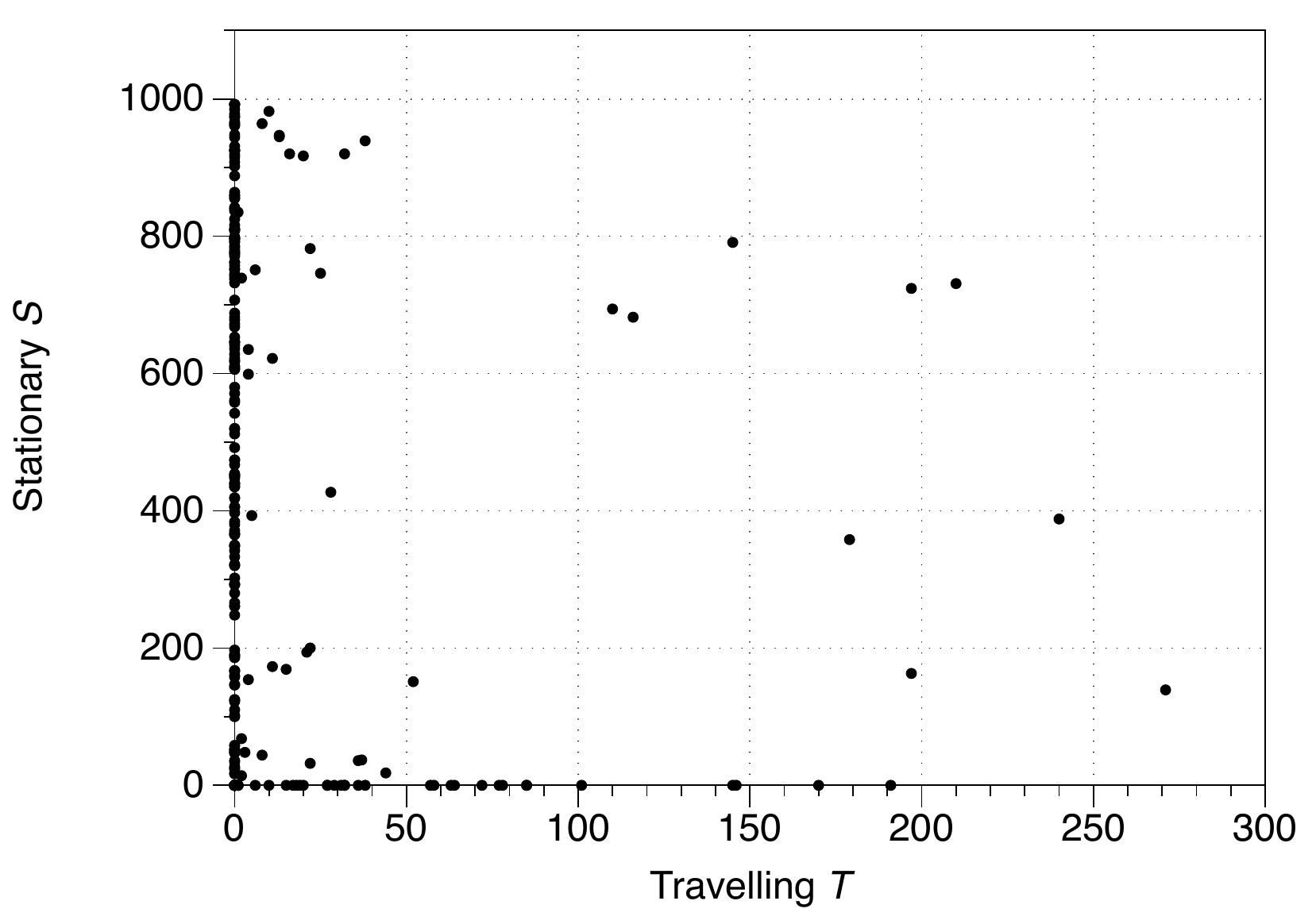}
\caption{Scatter of a  number $T$ of seeds supporting travelling localizations in each rule $R$ vs number $T$ of seeds supporting stationary localizations in the same rule $R$}
 \label{travellingvsstationary}
 \end{figure}
 
 \subsection{On rules supporting both stationary and travelling localisaitons}
 
 \begin{table}
\caption{Rules developing a maximum number of seeds into travelling and stationary localizations.}
\begin{tabular}{lll|lllllllll}
Rule & $F^0$ & $F^1$  & $T$ & $S$ & $H$  & $D$ &  $R$ & $P$ & $A$ & $I$ & $Z$ \\ \hline
(4,26) & 00100 & 11010 & 210 & 731 & 4.54 & 0.98	& 0.25 & 0.86 & 0.31 & 0.34 & 6.9E-6 \\
(5,25) & 00101 & 11001 & 197 & 724 & 3.83 & 0.95 & 0.25 & 0.73 & 0.14 & 0.19 & 1.3E-5 \\		
(7,4)   & 00111 & 00100 & 145 & 791 & 2.75 & 0.92 & 0.19 & 0.54 & 0.06 & 0.07 & 1.8E-4 \\
(5,26) & 00101 & 11010 & 240 & 388 & 4.70 & 0.99 & 0.25 & 0.92 & 0.39 & 0.39 & 6.4E-6 \\	
(6,21) & 00110 & 10101 & 110 & 694 & 4.40 & 0.98 & 0.25 & 0.70 & 0.14 & 0.14 & 1.3E-5 \\	
(7,5)  &  00111 & 00101 & 116 & 682 & 4.31 & 0.98 & 0.21 & 0.93 & 0.29 & 0.24 & 8.7E-6	\\
(7,21) & 00111 & 10101 & 179 & 358 & 4.31 & 0.98 & 0.22 & 0.97 & 0.56 & 0.39 &   6E-6	
\end{tabular}
\label{tablemaxlocal}
\end{table}

 Scatter of a  number $T$ of seeds supporting travelling localizations in each rule $R$ vs number $T$ of seeds supporting stationary localizations in the same rule $R$ is shown in Fig.~\ref{travellingvsstationary}. 
The plot shows that typically the more travelling localizations a rule exhibit the less stationary localizations the rule shows. There are however rules which  show high number of seeds leading to travelling and seeds leading to stationary localizations. They are shown in Tab.~\ref{tablemaxlocal}.

 \begin{figure}[!tbp]
\centering
\subfigure[]{\includegraphics[width=0.32\textwidth]{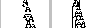}}
\subfigure[]{\includegraphics[width=0.32\textwidth]{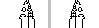}}
\subfigure[]{\includegraphics[width=0.32\textwidth]{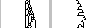}}
\subfigure[]{\includegraphics[width=0.32\textwidth]{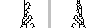}}
\subfigure[]{\includegraphics[width=0.32\textwidth]{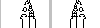}}
\subfigure[]{\includegraphics[width=0.32\textwidth]{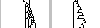}}
\subfigure[]{\includegraphics[width=0.32\textwidth]{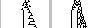}}
\subfigure[]{\includegraphics[width=0.32\textwidth]{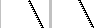}}
\subfigure[]{\includegraphics[width=0.32\textwidth]{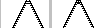}}
\subfigure[]{\includegraphics[width=0.32\textwidth]{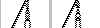}}
\subfigure[]{\includegraphics[width=0.32\textwidth]{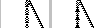}}
\subfigure[]{\includegraphics[width=0.32\textwidth]{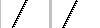}}
\subfigure[]{\includegraphics[width=0.32\textwidth]{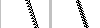}}
\subfigure[]{\includegraphics[width=0.32\textwidth]{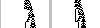}}
\subfigure[]{\includegraphics[width=0.32\textwidth]{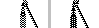}}
\subfigure[]{\includegraphics[width=0.32\textwidth]{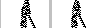}}
\subfigure[]{\includegraphics[width=0.32\textwidth]{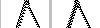}}
\caption{Exemplar configuration of rules from Tab.~\ref{tablemaxlocal}. In each subfigure space-time states of chain $x$, left, and $y$, right, are shown. 
(a) Rule (4, 26), seeds 01100 and 10001,
(b) Rule (5, 25), seeds   00100 and 00011,
(c) Rule (5, 25), seeds   00101 and 00011,
(d) Rule (5, 25), seeds  00101 and 01101,
(e) Rule (5, 25), seeds  00110 and 00001,
(f) Rule (5, 25), seeds 10001 and 10111,
(g) Rule (5, 25), seeds 11010 and 10011,
(h) Rule (7, 4), seeds 00001 and 00111,
(i) Rule (7, 4), seeds 00001 and 01111,
(j) Rule (7, 4), seeds 00001 and 11100,
(k) Rule (7, 5), seeds 00101 and 00111,
(l) Rule (7, 5), seeds 00111 and 00100,
(m) Rule (7, 21), seeds 00011 and 10110,
(n) Rule (7, 21), seeds 00011 and 11111,
(o) Rule (7, 21), seeds 00100 and 10001,
(p) Rule (7, 21), seeds 01001 and 01000,
(r) Rule (7, 21), seeds 10110 and 11100.
}
\label{exemplarconfigurations1}
\end{figure}

Configurations generated by rules in Tab.~\ref{tablemaxlocal}  from selected seeds  are shown in Fig.~\ref{exemplarconfigurations1}. 
Rules (4,26) and  (5,25) show a rich dynamics of traveling localizations (Fig.~\ref{exemplarconfigurations1}a--g). For most seeds the developments leads to formation of several propagating localizations, some of them could be classed as glider guns, often travelling localizations  produce stationary localizations during their interactions  with each other, e.g. Fig.~\ref{exemplarconfigurations1}be. In some cases,  e.g. Fig.~\ref{exemplarconfigurations1}cfg  
we can observe very elaborate trajectories of travelling licalisaitons.

In rule (7,4) a seed typically leads to formation of a single glider (Fig.~\ref{exemplarconfigurations1}h) or a pair of gliders moving in opposite directions (Fig.~\ref{exemplarconfigurations1}i); some seeds can produce a glider and
a stationary localisation (Fig.~\ref{exemplarconfigurations1}j).
Rule (7,5) exhibits combinations of travelling localizations and breathers: periodically oscillating stationary 
localizations (Fig.~\ref{exemplarconfigurations1}kl). 
 
A good variety of travelling and stationary localizations is given in space-time dynamics of rule (7,21). These
include large gliders propagating with a speed less than sped of light (`speed of light' glider translate one node per iteration) (Fig.~\ref{exemplarconfigurations1}m),  very slowly propagating localizations, or complex clusters of gliders, 
emitting stationary and travelling localisaitons (Fig.~\ref{exemplarconfigurations1}n),  combinations of large still lifes and gliders (Fig.~\ref{exemplarconfigurations1}p) and generation of two gliders of different sizes from a single seed (Fig.~\ref{exemplarconfigurations1}q).

 \begin{figure}[!tbp]
 \centering
  \includegraphics[width=0.49\textwidth]{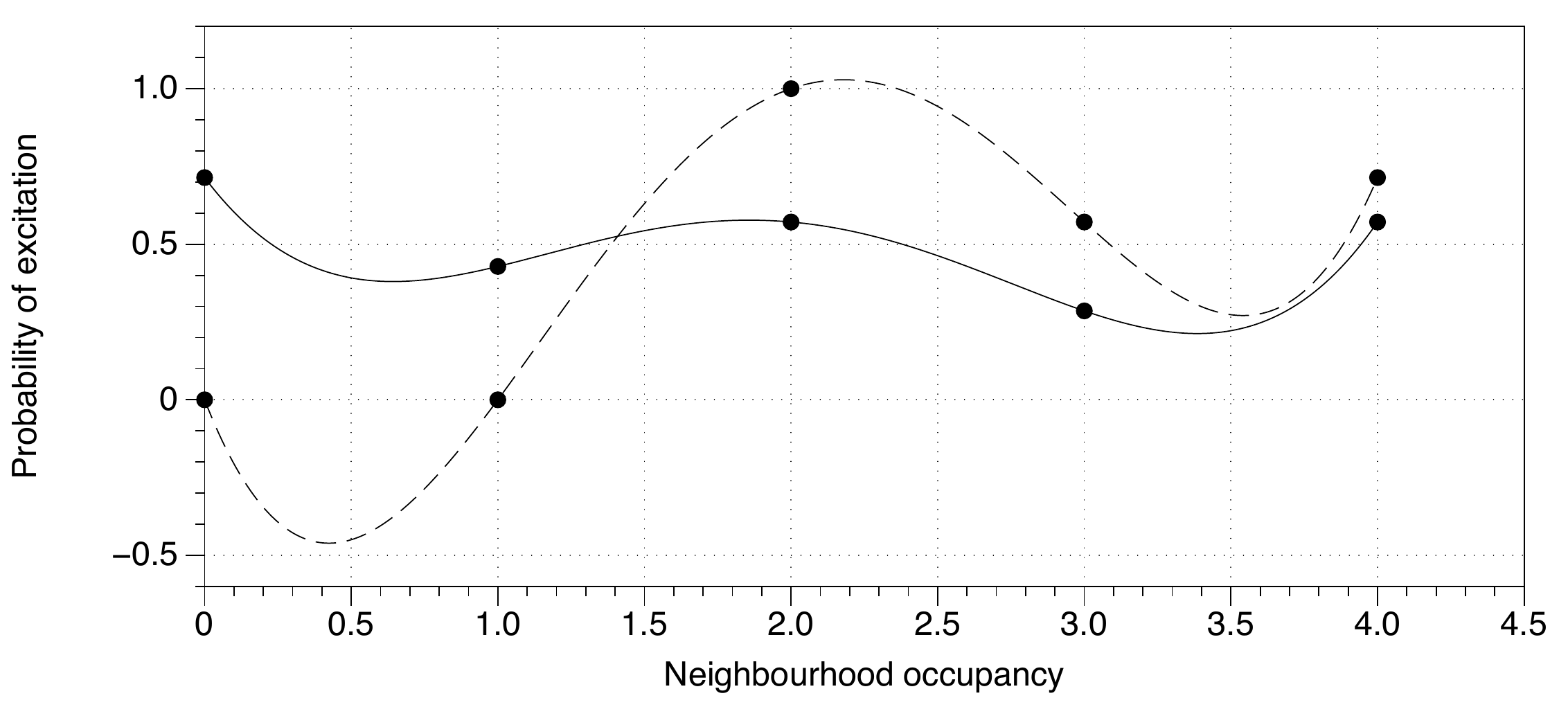}
  \caption{Dependence of a probability of excitation of a node in actin automaton supporting stationary and 
  travelling  localizations n a number of excited neighbours of the node. The polynomial of the probability is calculated on 7 rules that exhibit stationary and travelling  localizations for largest number of  seeds.   Dashed line shows probability of excitation of a resting node, solid line of an excited node, i.e. of an excited node to remain excited.}
 \label{travellingvsstationarypolynomials}
 \end{figure}

What is important for a rule to be supportive for both travelling and stationary localizations?  

Let us calculate frequency vectors $V^0$ and $V^1$ from $F^0$ and $F^1$. There are seven rules in Tab.~\ref{tablemaxlocal},  $V^0_i$ is a ratio of `1' in $F^0_i$ to 7, the same for $V_1$. 
 Thus we have $V^0 = (0 0 1 \frac{4}{7} \frac{5}{7})$ and $V^1 = (\frac{5}{7} \frac{3}{7} \frac{4}{7} \frac{2}{7} \frac{4}{7})$. Probabilities of excitation  can be represented via a polynomial approximation of vectors $V^0$ and $V^1$ shown in Fig.~\ref{travellingvsstationarypolynomials}. Assuming $0.5$ is a cut off frequency, we can transform the vector to a simplified form:  $^*V^0 = (0 0 1 1 1)$ and $^*V^1 = (1 0 1 0 1)$.
 
 \begin{proposition}
An actin automaton supports highest number of travelling and stationary localizations if its local activity is governed as follows. A resting node excites  if it necessarily has two excited neighbours or likely 
 three or more likely five excited neighbours.  An excited node remains excited if it has no excited neighbours or two or four excited neighbours.
 \end{proposition}
 
Thus rule $(7,4)$ is most representative in terms $F_0$ and rule $(6,12)$ is most representative 
 in terms of $F_1$. 

What integral characteristics are typical for the above localisation supporting rules?

The Shannon entropy $H$ is over 4 for most rules in Tab.~\ref{tablemaxlocal} but rule $(7,4)$. Simpson diversity index $D$ is over 0.9. Thus the rules which support both stationary and travelling localizations show very rich space time dynamics of excitations, $R$, when perturbed with a random initial pattern. They also show a moderate degree of space filling $D$, where, in case of random stimulation, 
 the dynamical excitation fills over 50\% of the automaton chains.  Rule (7,21) is a highest space filler, with almost 
 97\% of nodes being excited and rule (7,4) is a lowest space filler with just a half of nodes typically occupied by 
 excitation during the automaton development.  
 
 The space-time configurations are not overly rich morphologically with only circa 25\% of possible local configurations presented. Activity levels vary substantially between the rules in Tab.~\ref{tablemaxlocal}. Rule (7,4) shows lowest level of overall excitation activity followed by rules (5, 25) and (6,21). Highest level of activity are observed in space-time
 configurations of rule (7,21). Rules (7,4), (5,21) and (6,21) show highest compressibility $Z$, which is due to lower richness $R$ and, partly, space filling $P$. 
 
 \begin{proposition}
 Rules supporting both travelling and stationary localizations are characterised by high Shannon entropy and Simpson diversity, low levels of activity, and high degree of compressibility. 
 \end{proposition}
 
 \begin{table}
 \caption{Statistical characterisation of main groups of rules. 
 For each characteristics the table shows average and standard deviation}
 \begin{scriptsize}
 \begin{tabular}{l|llllll}
 	&	$\overline{H}, \sigma$	&	$\overline{D}, \sigma$	&	$\overline{R}, \sigma$	&	$\overline{P}, \sigma$	&	$\overline{A}, \sigma$	&	$\overline{I}, \sigma$	\\ \hline
All rules 	&	2.8, 1.2	&	0.8, 0.2	&	0.1, 0.1	&	0.9, 0.2	&	0.5, 0.2	&	0.2, 0.2	\\
Rules supporting travelling localizations	&	3.6, 0.9	&	0.9, 0.1	&	0.2, 0.0	&	0.8, 02	&	0.3, 0.2	&	0.3, 0.2	\\
Rules supporting stationary localizations 	&	2.3, 0.6	&	0.8, 0.1	&	0.1, 0.1	&	0.7, 0.2	&	0.1, 0.1	&	0.1, 0.0	\\
Rules supporting travelling and stationary localizations	&	4.1, 0.7	&	1.0, 0.0	&	0.2, 0.0	&	0.8, 0.2	&	0.3, 0.2	&	0.2, 0.1	\\
Rules supporting no localizations	&	2.9, 1.3	&	0.8, 0.2	&	0.2, 0.1	&	0.9, 0.1	&	0.6, 0.2	&	0.3, 0.2	\\
\end{tabular}
\end{scriptsize}
 \label{statistics}
 \end{table}
 
 Can we detect rules supporting localizations from integral characteristics of the space-time configurations generated
 by the rules?  Let us look at the Tab.~\ref{statistics}. We see that space filling $P$ is not a reliable discriminator but Shannon entropy $H$, activity level $A$ and incoherence $I$ are good discriminators on absence/presence and mobility of localizations. 
 
 \begin{proposition}
 Actin automata rules supporting highest number travelling localizations show high Shannon entropy $H$, 
 medium activity $A$ and high incoherence $I$. The rules supporting highest number of stationary localizations
 show low entropy $H$, low activity $A$ and low incoherence $I$. 
 \end{proposition}

\section{Discussion}

In an  exhaustive computational 
analysis of automaton models of hypothetical actin fibre communication events, we have discovered how to pinpoint rules supporting travelling and stationary 
localizations from global characteristics of space-time configurations generated by the actin automata rules. 
Nearly one third of node-state transition rules support either stationary or travelling localizations. Most the rules supporting localizations are `specialized': they support either stationary localizations or travelling localizations. Around 3\% of the rules shows rich dynamics of both travelling and stationary localizations, these rules are rare finds in the space of actin automaton node-state transitions.  
Rules supporting large numbers of travelling localizations usually produce space-time configurations with high Shannon entropy values, i.e. a higher potential information content. Rules which support both travelling and stationary localizations generate
configurations with high Shannon entropy and Simpson diversity index yet showing low levels of excitation activity 
and high degrees of compressibility. The rules supporting only stationary localizations exhibit low levels of 
excitation activity, low incoherence and low Shannon entropy. 

The above findings are valuable yet not efficient when searching for localizations in chains of biopolymers such as actin. It would seem to be more practical to decide on whether a rule supports localisation incidence by analysing the structure of the rule; we have proposed how this can be done. 

We also found that a rule supports travelling localizations if a resting node excites if it has two or four excited 
neighbours (50\% or 100\% of neighbourhood excitation) and an excited node remains excited if it has no excited neighbours. Indeed, additional modes of excitability might be necessary to support propagation of the localisation. A rule supports stationary localizations if a resting node excites if it has three excited neighbours (c. 70\% of neighbourhood is excited) and an excited node remains excited if it has less than four excited neighbours. 

But how can these models be applied to the proposed biophysical model of actin-based cellular communications? Whilst analysis of automaton patterns is inherently subjective \cite{Israeli_Goldenfeld_2006}, rules which support travelling localizations arguably form random patterns, whereas the few rules which support travelling and stationary localizations tend to appear more oscillatory. The observation the former, which may be described as emergent (i.e. apparent complexity arising from simple inputs and rules), provides a theoretical basis for describing the evolution and dynamics of complex patterns in biological systems. For example, if the actin cytoskeleton is considered as a sensorimotor data network, as we are suggesting, it may reasonably be hypothesised that emergent data patterns therein are a viable origin for the evolution of diverse, complex behaviours, even in unicellular organisms: such behaviours may be interpreted as intelligence, or at least efficient adaptations to cellular computation which favour survival. This finding substantiates our hypotheses in a previous paper \cite{Mayne_etal_2014}, in which we attempt to explain how apparently `intelligent' behaviour in uni/acellular organisms such as protists may be generated in a cytoskeletal cellular communications system.  

Our further studies will concern with developing working computing circuits on actin filament networks which topologies are closely matching topologies of the real intracellular networks. 


\end{document}